\documentclass[journal=nalefd,manuscript=article,layout=traditional]{achemso}

\usepackage{graphicx}
\usepackage{epsfig}
\usepackage{color}
\usepackage{amssymb}
\usepackage{amsmath}
\usepackage{array}
\usepackage[loose]{units}
\usepackage{xr-hyper}
\usepackage{hyperref}
\usepackage{cleveref}
\usepackage{braket}
\usepackage[caption=false]{subfig}
\usepackage[english]{babel}
\usepackage[T1]{fontenc}
\usepackage{letltxmacro}

\LetLtxMacro{\ORIGselectlanguage}{\selectlanguage}
\makeatletter
\DeclareRobustCommand{\selectlanguage}[1]{%
    \@ifundefined{alias@\string#1}
      {\ORIGselectlanguage{#1}}
      {\begingroup\edef\x{\endgroup
         \noexpand\ORIGselectlanguage{\@nameuse{alias@#1}}}\x}%
}
\newcommand{\definelanguagealias}[2]{%
  \@namedef{alias@#1}{#2}%
}
\makeatother

\definelanguagealias{en}{english}

\def\wph{\omega_{\rm ph}}      

\newcommand{\GOE}{Georg-August-Universit\"{a}t G\"{o}ttingen, D-37077 G\"{o}ttingen, Germany}
\newcommand{\ICFO}{ICFO-Institut de Ciencies Fotoniques, The Barcelona Institute of Science and Technology, 08860 Castelldefels (Barcelona), Spain}
\newcommand{\ICREA}{ICREA-Instituci\'{o} Catalana de Recerca i Estudis Avan\c{c}ats, Passeig Llu\'{i}s Companys 23, 08010 Barcelona, Spain}

\title{Probing chirality with inelastic electron-light scattering}
\author{Tyler R. Harvey}
\email{harvey@ph4.physik.uni-goettingen.de}
\affiliation{\GOE}
\author{Jan-Wilke Henke}
\affiliation{\GOE}
\author{Ofer Kfir}
\affiliation{\GOE}
\author{Hugo Loren\c{c}o-Martins}
\affiliation{\GOE}
\author{Armin Feist}
\affiliation{\GOE}
\author{F. Javier Garc\'{\i}a de Abajo}
\affiliation{\ICFO}
\alsoaffiliation{\ICREA}
\author{Claus Ropers}
\email{claus.ropers@uni-goettingen.de}
\affiliation{\GOE}
\date{\today}

\begin{document}
                            
\begin{abstract}
  Circular dichroism spectroscopy is an essential technique for understanding molecular structure and magnetic materials, but spatial resolution is limited by the wavelength of light, and sensitivity sufficient for single-molecule spectroscopy is challenging. We demonstrate that electrons can efficiently measure the interaction between circularly polarized light and chiral materials with deeply sub-wavelength resolution. By scanning a nanometer-sized focused electron beam across an optically-excited chiral nanostructure and measuring the electron energy spectrum at each probe position, we produce a high-spatial-resolution map of near-field dichroism. This technique offers a nanoscale view of a fundamental symmetry and could be employed as ``photon staining'' to increase biomolecular material contrast in electron microscopy. 
\end{abstract}

\maketitle

Chirality, defined as the absence of symmetry under spatial inversion, is central to a number of open scientific questions and technologically relevant materials, including the biochemistry of life \cite{evans_chirality_2012}, the weak force and its potential connection to the chirality of biomolecules \cite{dreiling_chirally_2014}, magnetic skyrmions \cite{yu_real-space_2010}, and metamaterials \cite{tretyakov_magnetoelectric_1998,zhang_negative_2009}. Tools to probe chirality allow us insight into these phenomena; the first structural information about organic compounds came through studying the rotation of linearly polarized light in glucose solutions \cite{drayer_early_2001}. The Kramers-Kronig-related technique, circular dichroism (CD), is based on differences in absorption or scattering of circularly polarized light. 

CD spectroscopy is widely employed to understand microscopic structure. X-ray magnetic circular dichroism \cite{thole_x-ray_1992,stohr_x-ray_1995} is a routine method for nanoscale magnetic domain imaging \cite{stohr_principles_1998}. In biochemistry, circular dichroism spectroscopy serves as a molecular structural fingerprinting technique: the combination of local structural chirality and an optical transition necessary to produce a peak in a CD spectrum is more selective than just an optical transition \cite{purdie_analytical_1989}. Distinct kinds of molecular structural information are accessible through circular dichroism depending on the wavelength range. Secondary structure, e.g. $\alpha$-helices and $\beta$-sheets, small collections of hydrogen bonds that constitute segments of proteins, produce distinguishable far-UV CD spectra \cite{johnson_protein_1990}. Near-UV CD spectra are sensitive to tertiary structure, the global arrangement of all of the segments \cite{li_applications_2011}. Vibrational circular dichroism can retrieve structural information in conjunction with DFT calculations \cite{stephens_determination_2008}.

These far-field circular dichroism techniques have some limitations. Because of the diffraction limit, spatial resolution is tied to the wavelength of light. The chiral contribution to molecular absorption is many orders of magnitude smaller than the non-chiral contribution; limited sensitivity necessitates the use of large ensembles of molecules \cite{purdie_analytical_1989,rhee_amplifications_2013}. One potential route toward higher sensitivity is to enhance the chiral part of optical absorption with spatially-structured fields \cite{tang_optical_2010,bliokh_characterizing_2011,choi_limitations_2012,coles_chirality_2012,rhee_amplifications_2013,barr_investigating_2018}. Another option is to measure near fields. Scanning near-field optical microscopy can be employed to measure near-field circular dichroism at surfaces, but resolution and bandwidth are limited by the tip, and care must be taken to preserve polarization \cite{narushima_circular_2013,schnell_real-space_2016}. Photoemission electron microscopy can measure optical near fields with high spatial and temporal resolution \cite{kubo_femtosecond_2007,kahl_normal-incidence_2014,gong_ultrafast_2015,el-khoury_visualizing_2016,dabrowski_ultrafast_2017,frank_short-range_2017} and illuminate spin-dependent plasmon dynamics \cite{spektor_revealing_2017,dai_ultrafast_2018,spektor_mixing_2019}; the technique is also commonly used in conjunction with x-ray magnetic circular dichroism \cite{vogel_time-resolved_2003,cheng_studies_2012}. Cathodoluminescence polarimetry is a promising new approach to probe optical interactions with nanometer resolution and access to polarization \cite{osorio_angle-resolved_2016}. Dichroism via exchange of electron orbital angular momentum and energy offers similar information without light \cite{asenjo-garcia_dichroism_2014,harvey_demonstration_2015,guzzinati_probing_2017,zanfrognini_orbital_2019}.

\begin{figure}[h]
  \subfloat[]{\label{subfig:setup:cartoon}
  \includegraphics[height=1.5in]{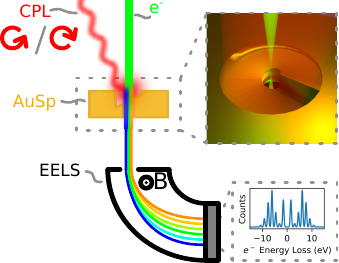}}
  \subfloat[]{\label{subfig:setup:specimen}
  \includegraphics[height=1.45in]{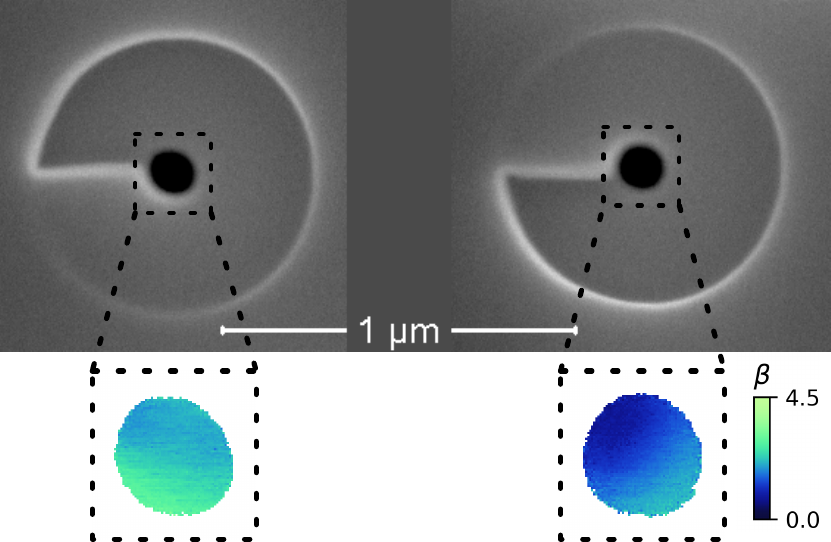}}
  \caption{(a) Schematic of the experiment. The probe electron pulse ($e^-$, green) and circularly polarized optical pulse (CPL, red) are nearly colinearly incident on the specimen (AuSp, gold). Coupling to the optical field produces electron energy sidebands that are measured in an electron energy-loss spectrometer (EELS, black). Measurement of this coupling, $|\beta_{\sigma}|(\mathbf{r}_{\perp})$, at each electron probe position produces a spatial map. (b) Scanning electron micrograph of the specimen measured in Fig. \ref{fig:beta_maps}, $\unit[400]{\textrm{nm}}$ pitch, $\unit[800]{\textrm{nm}}$ outer diameter spirals of opposite handedness milled into a $\unit[1]{\mu m}$-thick gold slab with a $\unit[100]{\textrm{nm}}$ diameter hole. Insets show PINEM maps with left-circularly polarized light, as shown in Fig. \ref{subfig:B:expt}. }
\end{figure}

Measurement of externally-pumped optical near fields is possible in transmission electron microscopes. Although free-space absorption or emission of a photon by an electron is forbidden as it cannot conserve both energy and momentum, electrons in optical fields adjacent to or inside materials can exchange quantized momentum and energy with the field \cite{saathoff_laser-assisted_2008,garcia_de_abajo_electron_2008}. Barwick {\it et al.} employed this interaction to map optical near fields driven by laser illumination of the specimen, called photon-induced near-field electron microscopy (PINEM) \cite{barwick_photon-induced_2009}. As near fields in an ultrafast transmission electron microsope can be sufficiently strong that the probability for an electron to lose or gain one or more units of photon energy is higher than the probability to exchange no energy \cite{park_photon-induced_2010,garcia_de_abajo_multiphoton_2010,feist_quantum_2015}, PINEM is a highly sensitive probe of optical near fields and has been employed to image single-atom-high step edges \cite{park_graphene-layered_2013}, proteins \cite{flannigan_biological_2010}, cells \cite{kaplan_photon-induced_2017}, and nanoparticles \cite{yurtsever_direct_2012}. While most applications of PINEM so far have relied on ultrashort electron and optical pulses to increase the strength of the optical field, the same interaction is possible with a continuous electron beam \cite{das_stimulated_2019,liu_continuous_2019}.
In this work, we demonstrate a near-field circular dichroism technique based on inelastic electron-light scattering. 
A unitless parameter $\beta$ characterizes the local strength of the coupling between the near field and the electron beam \cite{garcia_de_abajo_electron_2016,park_photon-induced_2010,garcia_de_abajo_multiphoton_2010}, where 
\begin{equation}  \label{eq:beta}
  \beta_{\sigma}(\mathbf{r}_{\perp}) = \frac{e}{\hbar \wph} \int \mathrm{d}z \ E_z(\mathbf{r}) e^{-i\wph z / ve},
\end{equation}
depends on transverse electron beam position $\mathbf{r}_{\perp}$, electron velocity $ve$, and the longitudinal component of the local electric field in or near the material $E_z(\mathbf{r},\sigma,\wph)$. This field depends on optical illumination helicity $\sigma$ and photon energy $\hbar \wph$. The exchange of $m$ quanta of energy with the optical field produces an electron energy distribution with a series of peaks at energies separated by $\hbar \wph$ with probabilities $P_m = |J_m(2|\beta|)|^2$ \cite{park_photon-induced_2010,garcia_de_abajo_multiphoton_2010,feist_quantum_2015} (see Supplemental page 5 for a more detailed description of the resulting spectra). A calculated example spectrum is shown in Fig. \ref{subfig:setup:cartoon}.

Because electron beams can be focused to sub-nanometer spots in modern electron microscopes, this interaction probes the optical response of materials with nanometer spatial resolution. 
Electron near-field circular dichroism (ENFCD) then can be defined as the normalized difference in $\beta$ measured with left- and right-circularly polarized light.
\begin{equation} \label{eq:Delta}
  \Delta(\mathbf{r}_{\perp}) = \frac{|\beta_{LCP}| - |\beta_{RCP}|}{|\beta_{LCP}|+|\beta_{RCP}|}
\end{equation}
Spatial variations in $\Delta$ offer insight into the microscopic origins of chiral optical responses. Access to this information is not available at high spatial resolution through far-field circular dichroism, where the response is typically modeled with a small number of dipole transition moments \cite{power_circular_1974}. This contrast mechanism can be employed for ``photon staining'' to boost contrast on biomolecular materials that only weakly scatter electrons.  

\begin{figure}[h]
  \subfloat[]{\label{subfig:B:model}
  \includegraphics[height=1.5in]{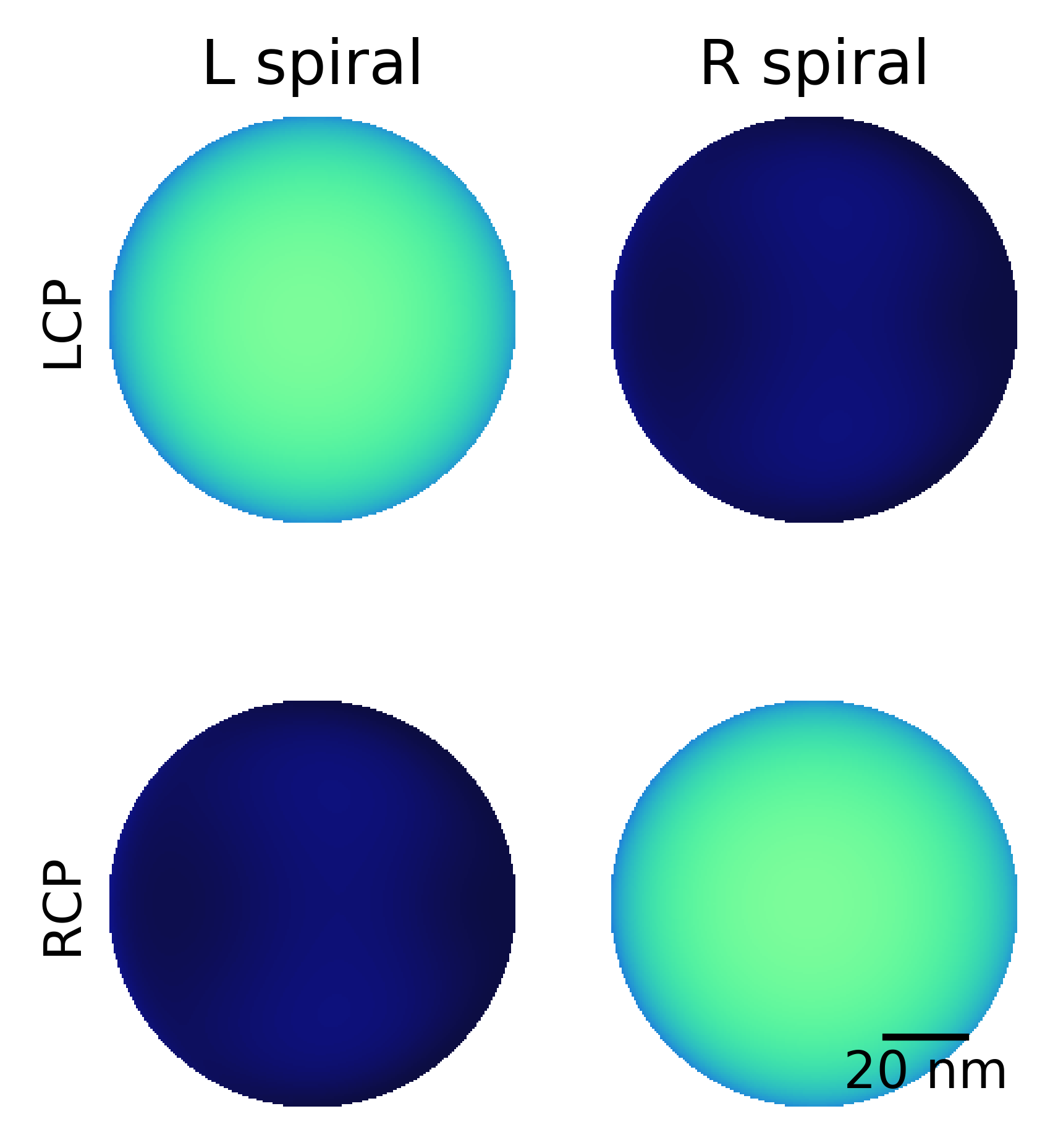}}
  \subfloat[]{\label{subfig:B:sim}
  \includegraphics[height=1.5in]{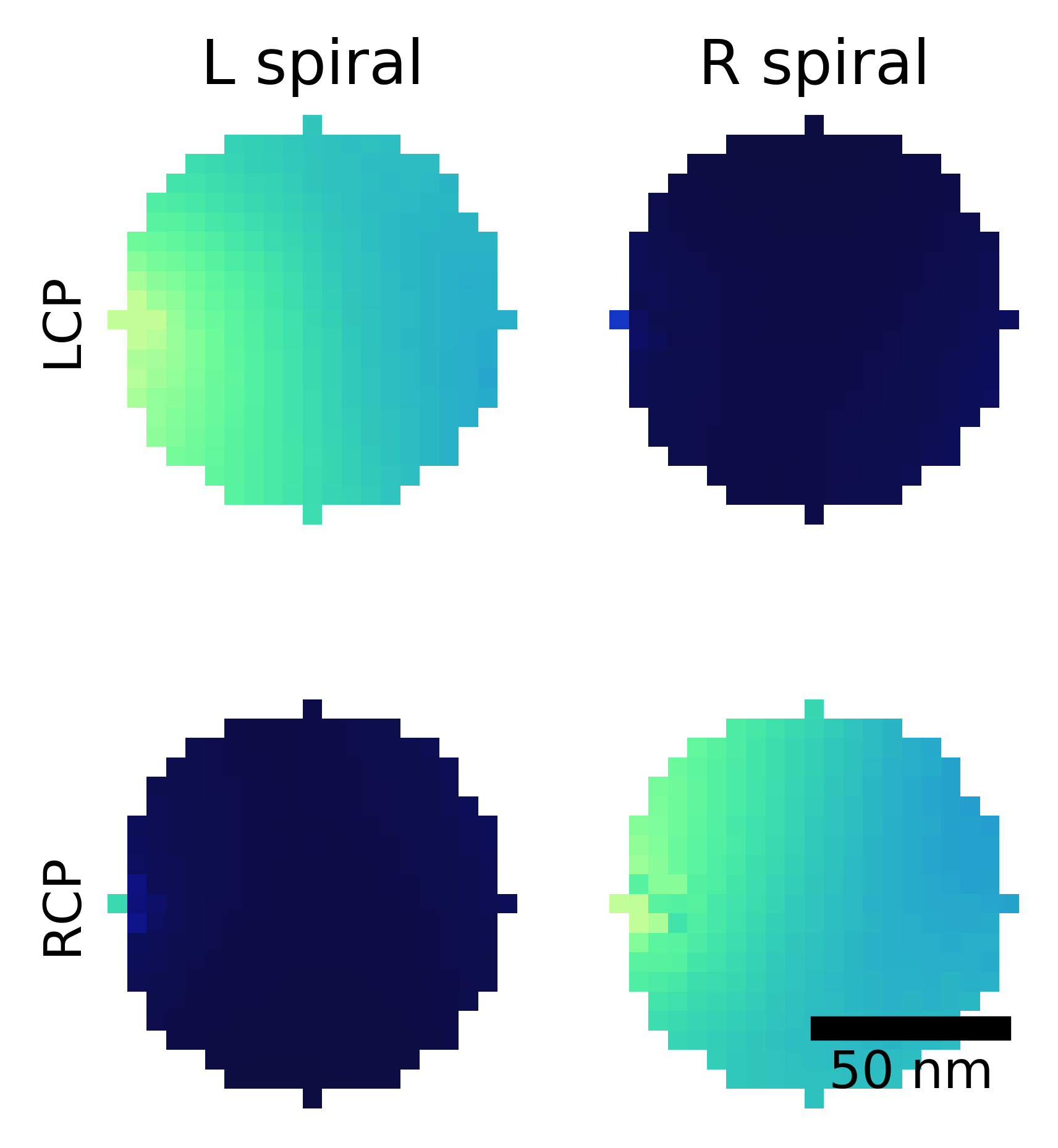}}
  \subfloat[]{\label{subfig:B:expt}
  \includegraphics[height=1.5in]{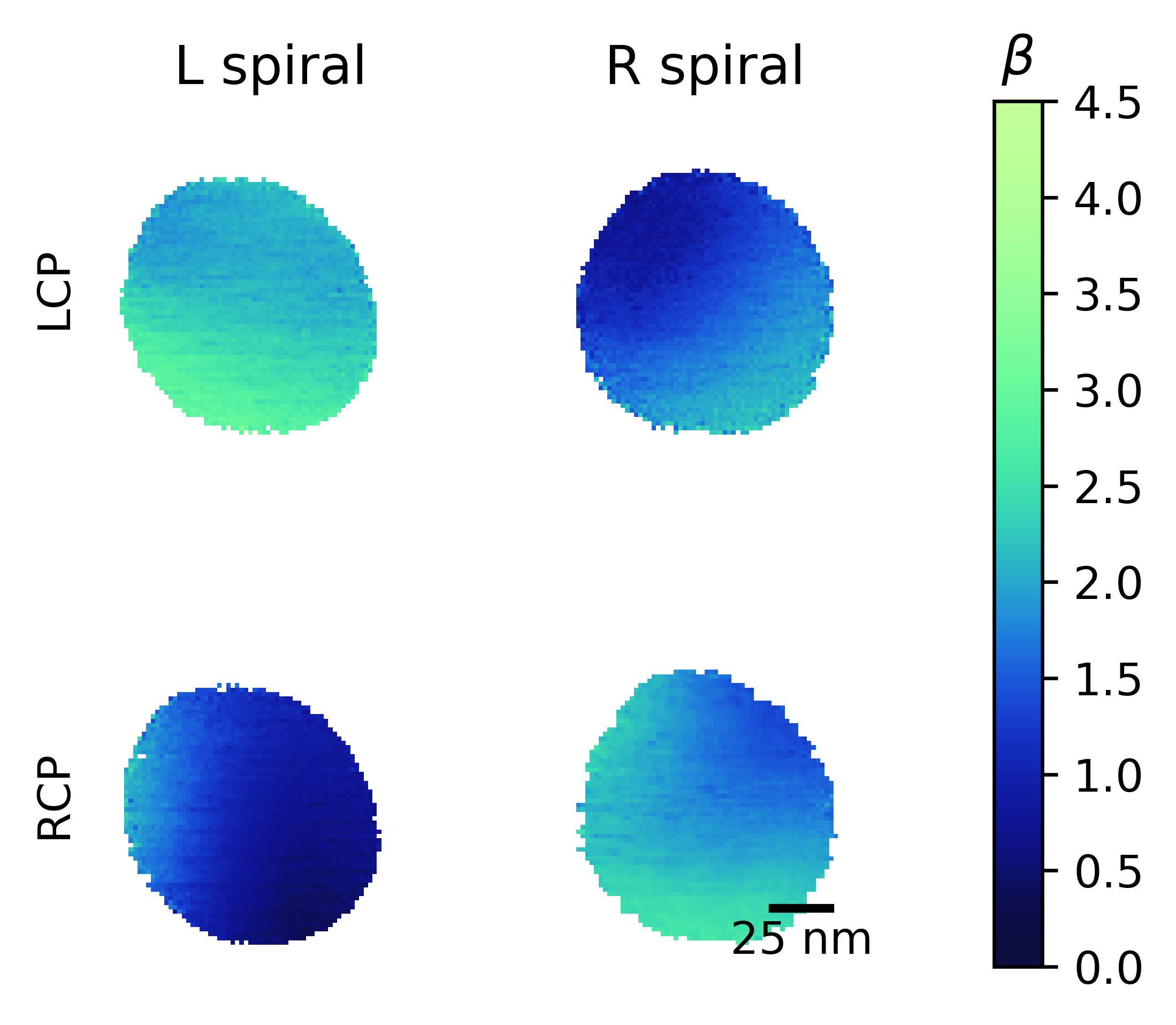}} \\
  \subfloat[]{\label{subfig:d:model}
  \includegraphics[height=0.8275in]{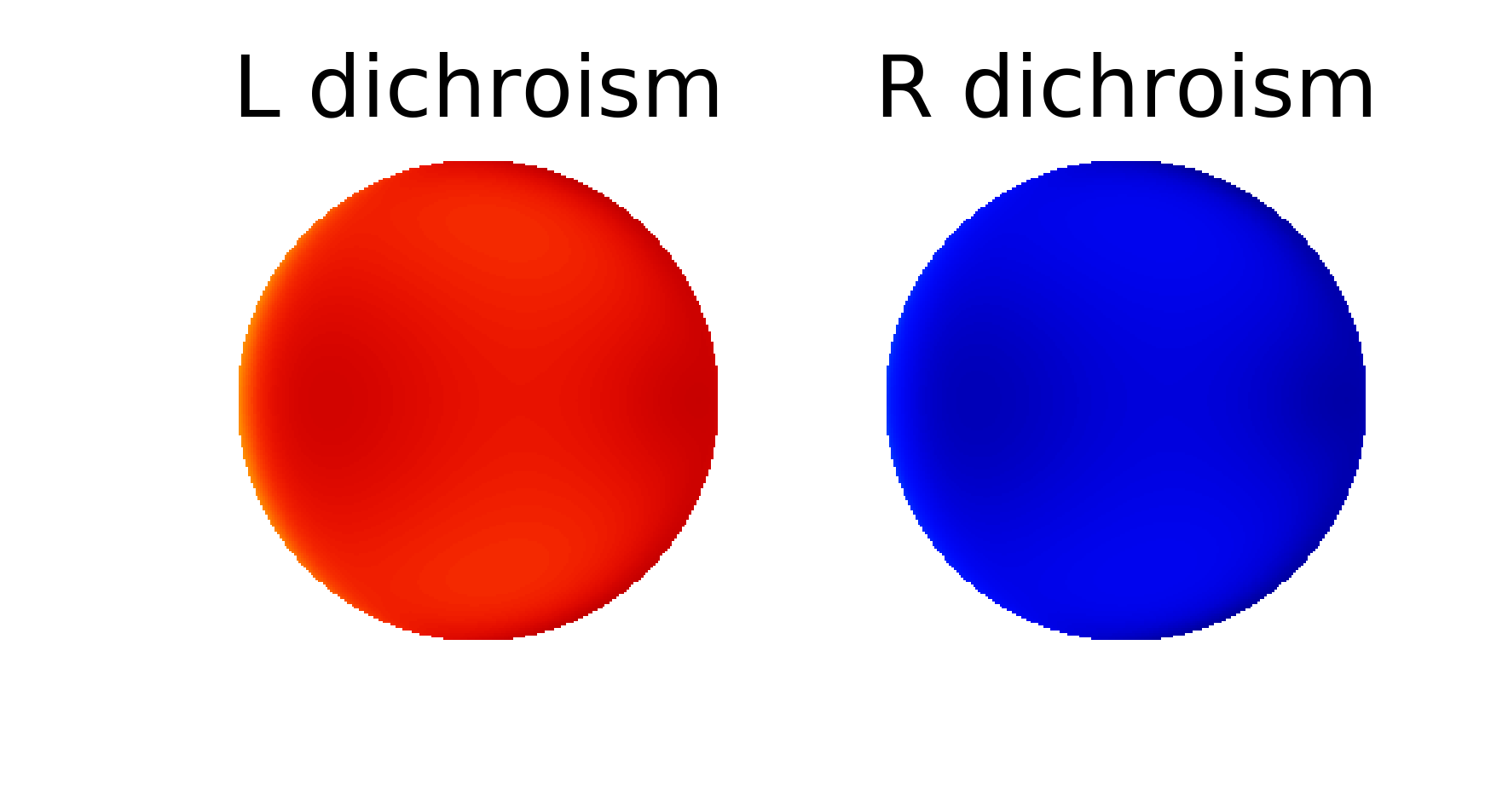}}
  \subfloat[]{\label{subfig:d:sim}
  \includegraphics[height=0.8275in]{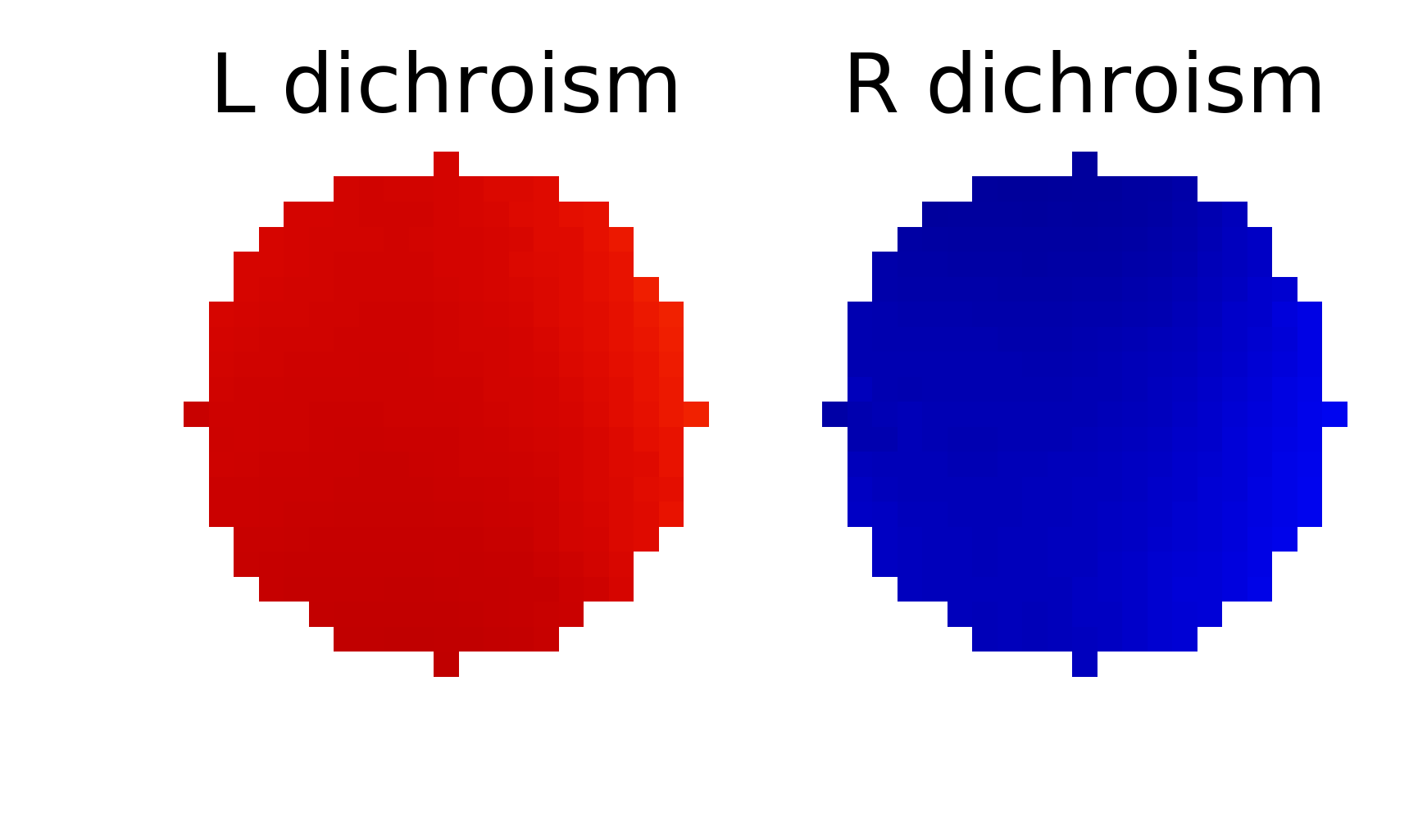}}
  \subfloat[]{\label{subfig:d:expt}
  \includegraphics[height=0.8275in]{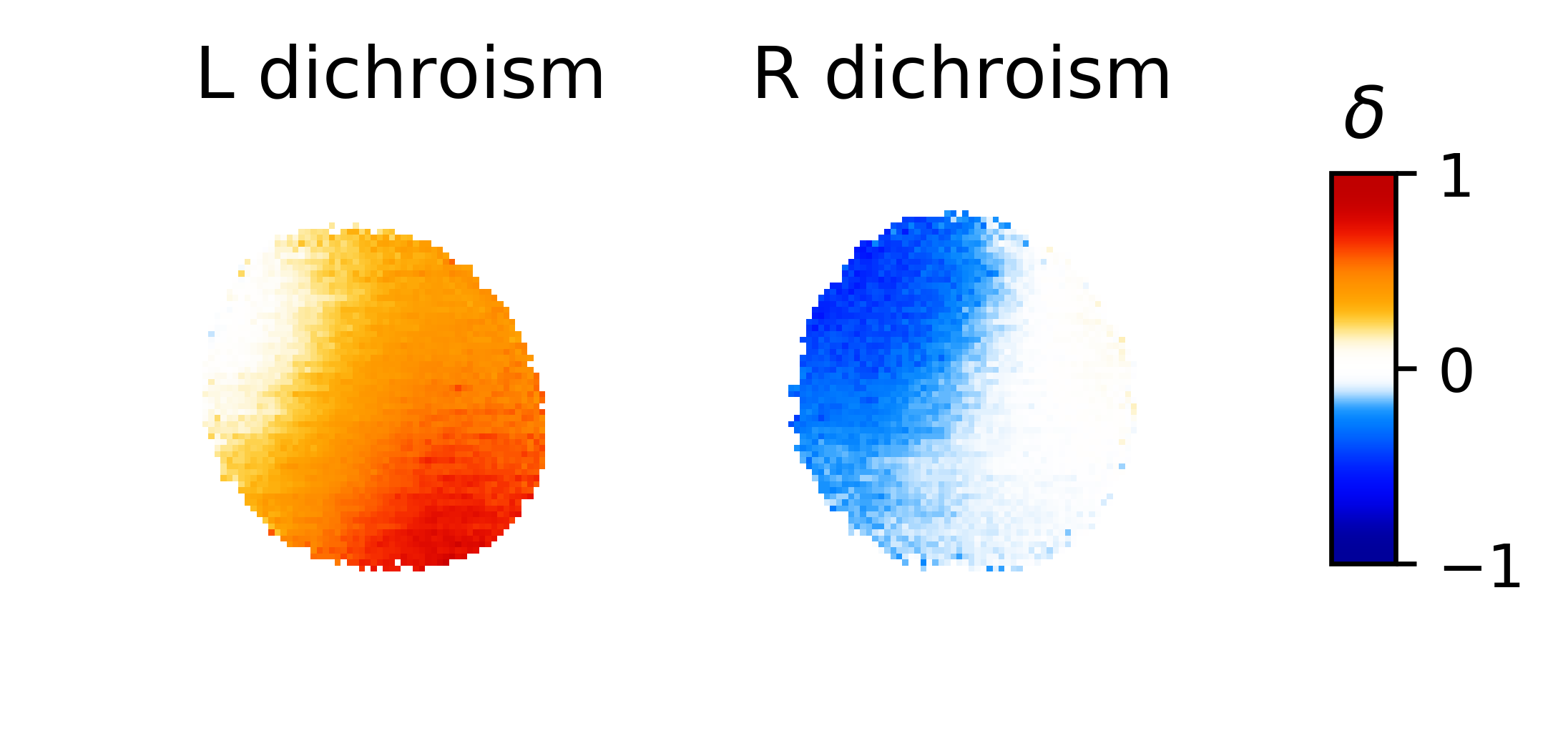}} \\
  \caption{(a) Modeled, (b) simulated and (c) experimental maps of the coupling constant $\beta_{\sigma}(\mathbf{r}_{\perp})$ across the hole in the left-handed structure (left) and the right-handed structure (right) with left-circularly-polarized (LCP, top) and right-circularly-polarized (RCP, bottom) light. (d) Modeled, (e) simulated and (f) experimental maps of the dichroism $\Delta(\mathbf{r}_{\perp})$. \label{fig:beta_maps}} 
\end{figure}

We measured ENFCD on a prototypical chiral specimen that exemplifies this non-dipolar spatial variation, as shown in Fig. \ref{subfig:setup:specimen}. We prepared two nearly identical spirals of opposite handedness by FIB milling a $\unit[1]{\mu m}$-thick gold slab. As the chirality in this specimen arises from geometry, we numerically calculated ENFCD in these structures based on frequency-domain finite element simulation of the optical fields with the COMSOL Multiphysics package. The simulations (Fig. \ref{subfig:B:sim}) show that the spatial average of $\beta$ inside the hole in the structure strongly depends on the optical helicity, and the relationship flips with the structure's handedness. In order to better understand the physical origins of spatial variations in ENFCD, we also developed a model for the optical response of the specimen, which we describe in more detail below and on Supplemental page 7. The model (Fig. \ref{subfig:B:model}) also shows a helicity-dependent average $\beta$ 

We performed these measurements using the G\"{o}ttingen JEOL 2100F ultrafast transmission electron microscope \cite{feist_ultrafast_2017}. Circularly polarized $\unit[5]{ps}$, $\unit[42]{nJ}$ optical pump pulses with $\unit[800]{nm}$ central wavelength illuminated the specimen at nearly normal incidence with a $\unit[600]{kHz}$ repetition rate, as shown in Fig \ref{subfig:setup:cartoon}. We raster-scanned $\unit[1]{ps}$ electron probe pulses focused to a spot size smaller than $\unit[10]{nm}$ over the hole in the spiral structures. An electron energy-loss spectrometer recorded an electron spectrum at each probe position, and we measured $\beta$ from each spectrum. The resulting experimental maps of $\beta_{\sigma}(\mathbf{r}_{\perp})$, shown in Fig. \ref{subfig:B:expt}, agree with simulations regarding the relationship between structure and illumination handedness but show more spatial variation than both the model (Fig. \ref{subfig:B:model}) and simulations (Fig. \ref{subfig:B:sim}). The larger spatial variation in the experimental maps likely arises due to additional contributions to the optical response from real features (e.g. smoother edges, finite slope of the spiral cut and small optical coupling with other features in the surrounding region) that are not included in the simulations. These nanoscale spatial variations in ENFCD offer information about three-dimensional structure and insight into the microscopic origins of chirality.

We considered two ways to interpet the spatial variations in ENFCD. First, for molecules or other specimens with a finite number of electronic transitions with limited coupling between them, it may be possible to match measured ENFCD maps with a model based on a sum of dipole transition moments. In the Supplemental Material (Equation 11), we calculated the PINEM maps expected in this case. Second, for specimens where collective excitations of many states contribute most to the optical response, we developed a model based on a simplified one-dimensional structure. At the $\unit[800]{\textrm{nm}}$ optical wavelength we used, gold can be treated as a good conductor with a skin depth much smaller than the size of the holes in the spirals we used. We therefore represent these spirals as a 1D single-turn helix with a magnetic polarizability along the axis of the helix, corresponding to a magnetic polarizability along the inner edge of the hole in the spiral.

\begin{figure}[h]
  \subfloat[]{\label{subfig:model:schematic}
  \includegraphics[height=1.5in]{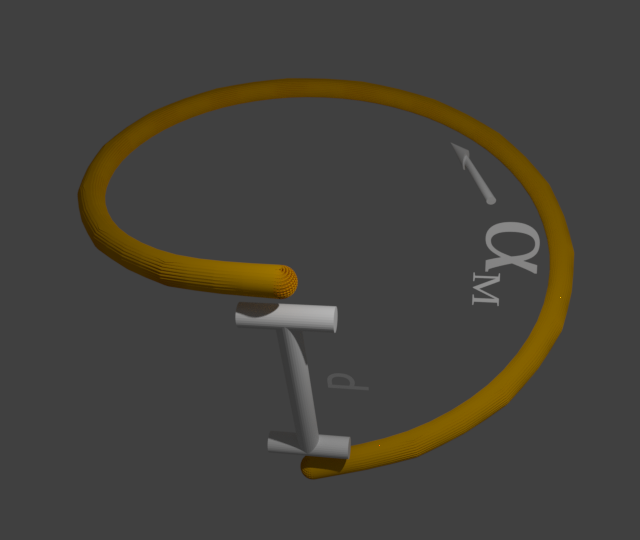}}
  \subfloat[]{\label{subfig:pitch:model}
  \includegraphics[height=1.5in]{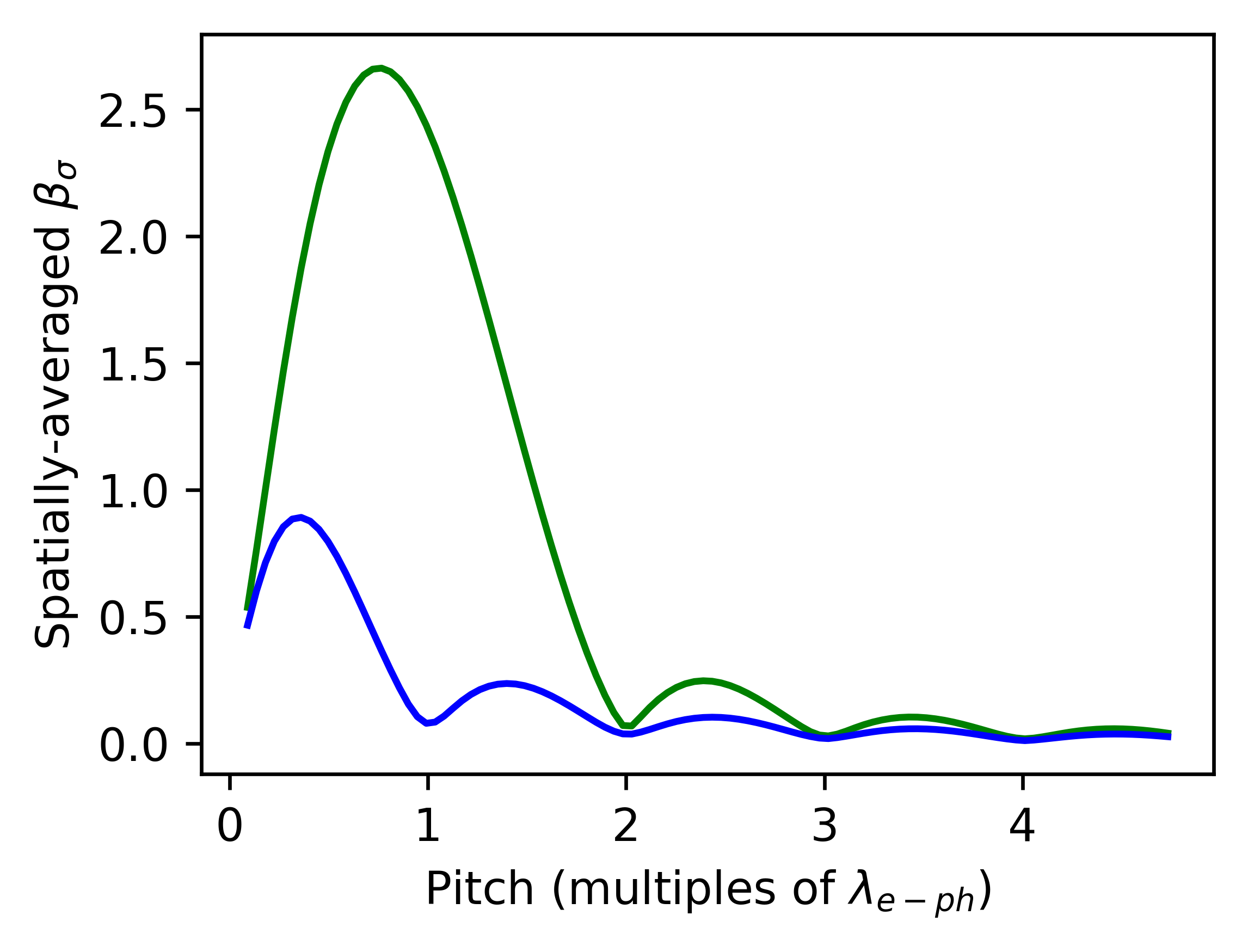}}
  \subfloat[]{\label{subfig:energy:both}
  \includegraphics[height=1.5in]{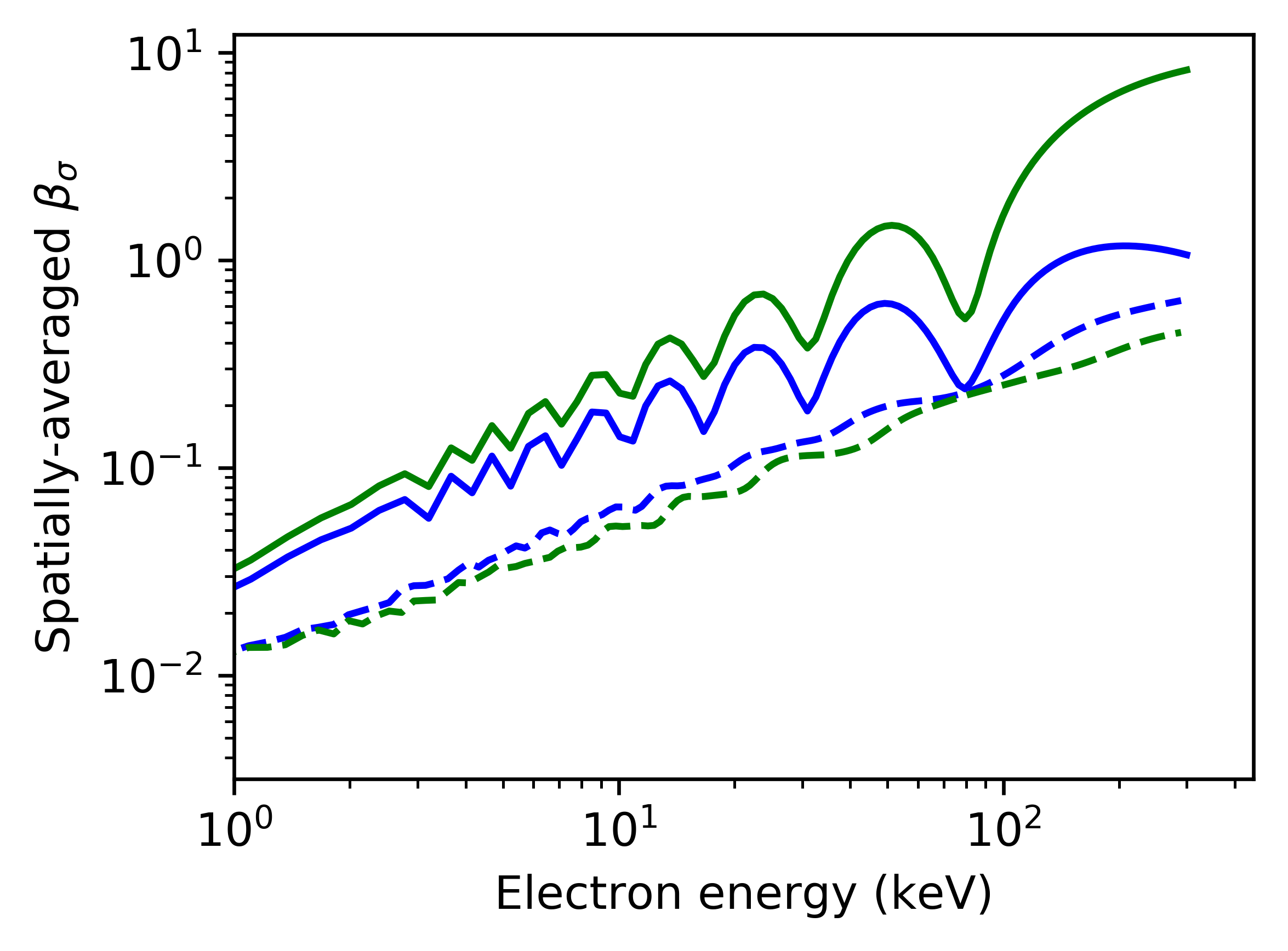}}
  \caption{(a) Schematic of the modeled 1D helix with magnetic polarizability $\alpha_M$ along the helix axis and pitch $d$. (b) Calculated dependence of the average coupling constants $\beta_{LCP}$ (blue) and $\beta_{RCP}$ (green) on the pitch of the model 1D helix with diameter $\unit[100]{\textrm{nm}}$, optical wavelength $\unit[800]{\textrm{nm}}$ and electron energy $\unit[200]{\textrm{keV}}$. The minima occur at integer multiples of the coupling period $\lambda_{e-ph} = \unit[554]{\textrm{nm}}$. (c) Calculated electron energy dependence of the average coupling constants from the model (solid lines) and from finite element simulations (dashed lines) with helix pitch and optical wavelength both $\unit[800]{\textrm{nm}}$. The same oscillatory behavior occurs as $\lambda_{e-ph}$ depends on energy, although it is only qualitatively similar behavior in the finite element simulation as there are likely other contributions to the response. Note that when the helix pitch matches the optical wavelength, the resonant coupling energy is infinity, so no maximum occurs here.}
\end{figure}

This model predicts an oscillatory response of the integrated coupling constant as a function of pitch $d$ of the spiral for both circular polarizations within the hole, as shown in Fig. \ref{subfig:pitch:model}, with a coupling period corresponding to the momentum mismatch $\lambda_{e-ph} = \frac{2\pi ve}{\wph}$. At $d = \lambda_{e-ph}$, the integrated relative dichroism is maximized. When the pitch is an integer multiple of $\lambda_{e-ph}$, the spatial symmetries of the coupling constant (Fig. 4 in the Supplemental) suggest that spin-to-orbital angular momentum conversion \cite{spektor_revealing_2017,dai_ultrafast_2018,spektor_mixing_2019,vanacore_ultrafast_2019} determines the response: at a pitch $ d = \ell \lambda_{e-ph}$, the coupling constant has $\ell+\sigma$ rotational symmetry, where $\sigma$ is the helicity of the exciting light.
The resonant dependence of the coupling constant on pitch as a function of $\lambda_{e-ph}$ suggests that we can boost the strength of dichroism, and therefore structural information obtained, by tuning either the electron kinetic energy or the optical wavelength. We have a recipe for maximization of the coupling constant and ENFCD for specimens with helical structural components such as $\alpha$-helices or chiral plasmonic structures \cite{fan_plasmonic_2010,kuzyk_DNA-based_2012,hentschel_three-dimensional_2012,shen_three-dimensional_2013,song_tailorable_2013,hentschel_chiral_2017,saito_chiral_2018}. Specifically, by first setting the optical wavelength to known optical resonances and then tuning the electron energy to match $\lambda_{e-ph}$ to the helical pitch $d$ of the structure. For example, for helical assemblies of semiconducting particles with a $\unit[390]{\textrm{nm}}$ pitch and a peak in optical circular dichroism at $\sim\unit[690]{\textrm{nm}}$ \cite{yan_self-assembly_2020}, resonant coupling occurs at $\unit[108]{\textrm{keV}}$. This resonant coupling and high spatial resolution could be employed to image the assembly of helical nanostructures \cite{srivastava_light-controlled_2010,yan_self-assembly_2012,kuzyk_DNA-based_2012,shen_three-dimensional_2013,song_tailorable_2013} {\it in situ}, or to visualize optical coupling between molecules and nanostructures for molecular sensing \cite{james_chiral_1995,garcia-guirado_enantiomer-selective_2018}.
It may be possible to produce stronger dichroism using these two knobs than is possible in a linear optical technique where the light wavelength at which absorption occurs is far from the characteristic length scales associated with chirality.

We have demonstrated a technique to measure circular dichroism in optical near fields with nanometer resolution. The spatially varying near-field dichroism we measure offers nanoscale detail not accessible through far-field optical circular dichroism. 
More information will be available with ENFCD spectroscopy, using a range of optical pump wavelengths to measure the full spatial and spectral response of a specimen. The available wavelength range is limited only by the laser source and optics used to illuminate the sample and can therefore be very broad. On the other hand, it may be possible to achieve single-molecule sensitivity over a limited wavelength range with the Purcell effect through coupling of molecular electronic states to the substrate or nanostructures. The local nature of our measurement means that coupling to an electric dipole is sufficient to enhance sensitivity \cite{garcia-etxarri_surface-enhanced_2013}. As inelastic electron-light scattering is coherent, further sensitivity through interferometric measurement of the coupling constant is possible \cite{echternkamp_ramsey-type_2016,priebe_attosecond_2017,madan_holographic_2019}.
Temporal resolution is also available with ENFCD. The probe pulses we used in this work always arrived at the peak of the optical pump pulses, but temporal measurement of long-lived optical excitations in chiral materials is straightforward by varying probe-pulse arrival time. Chiral magnetization, polarization or structural dynamics could be probed with ENFCD using a first optical pulse to pump the specimen and a second, weaker optical pulse overlapped with the delayed probe pulse. Access to nanoscale near-field circular dichroism spectra enables high-spatial-resolution characterization of three-dimensional structure with straightforward extension to temporal resolution and possibilities for single-molecule sensitivity.
  \section*{Acknowledgments}
  We appreciate the work the G\"{o}ttingen UTEM team has put into developing the microscope. We acknowledge helful discussions with Katherina Echternkamp, Murat Sivis, and Benjamin McMorran. We acknowledge financial support by the Deutsche Forschungsgemeinschaft through SPP 1840 QUTIF and the Leibniz program. O.K. gratefully acknowledges funding from the European Union’s Horizon 2020 research and innovation programme under the Marie Sk\l{}odowska-Curie grant agreement No.752533. F.J.G.A. acknowledges support from ERC (advanced grant 789104-eNANO) and the Spanish MINECO (MAT2017-88492- R and SEV2015-0522). T.R.H. acknowledges the support of a postdoctoral fellowship from Alexander von Humboldt Foundation and its sponsor, the German Federal Ministry for Education and Research.

\bibliography{cPINEM_paper}{}
\end{document}


\maketitle
\abstract{In the supplementary material, we show further measurements of electron near-field circular dichroism and linear dichroism measurements and detail our polarization calibration procedure. We describe the procedure we used in finite-element simulations. Finally, we calculate the coupling constant for inelastic electron-photon interactions with oscillating electric and magnetic dipoles, and then model our specimen with a 1D polarizability.}

\section{Additional Dichroism Measurements}

\begin{figure}
  \begin{minipage}{0.4\textwidth}
    \subfloat[]{\label{subfig:red:sem} 
  \includegraphics[height=0.5in]{red_0deg_cropped_brightened.png}} \\
  \subfloat[]{\label{subfig:red:beta}
  \includegraphics[height=1.5in]{red_04_g_w_cb.png}} \\
  \subfloat[]{\label{subfig:red:dich}
  \includegraphics[height=0.8275in]{red_04_norm_dichroism_w_cb.png}}
  \end{minipage}
  \begin{minipage}{0.4\textwidth}
    \subfloat[]{\label{subfig:forest:sem} 
  \includegraphics[height=0.5in]{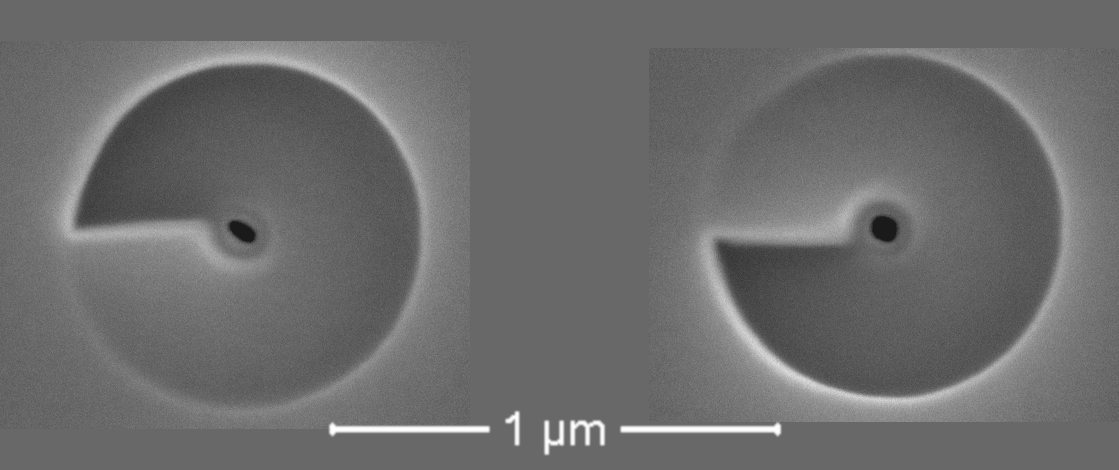}} \\
  \subfloat[]{\label{subfig:forest:beta}
  \includegraphics[height=1.5in]{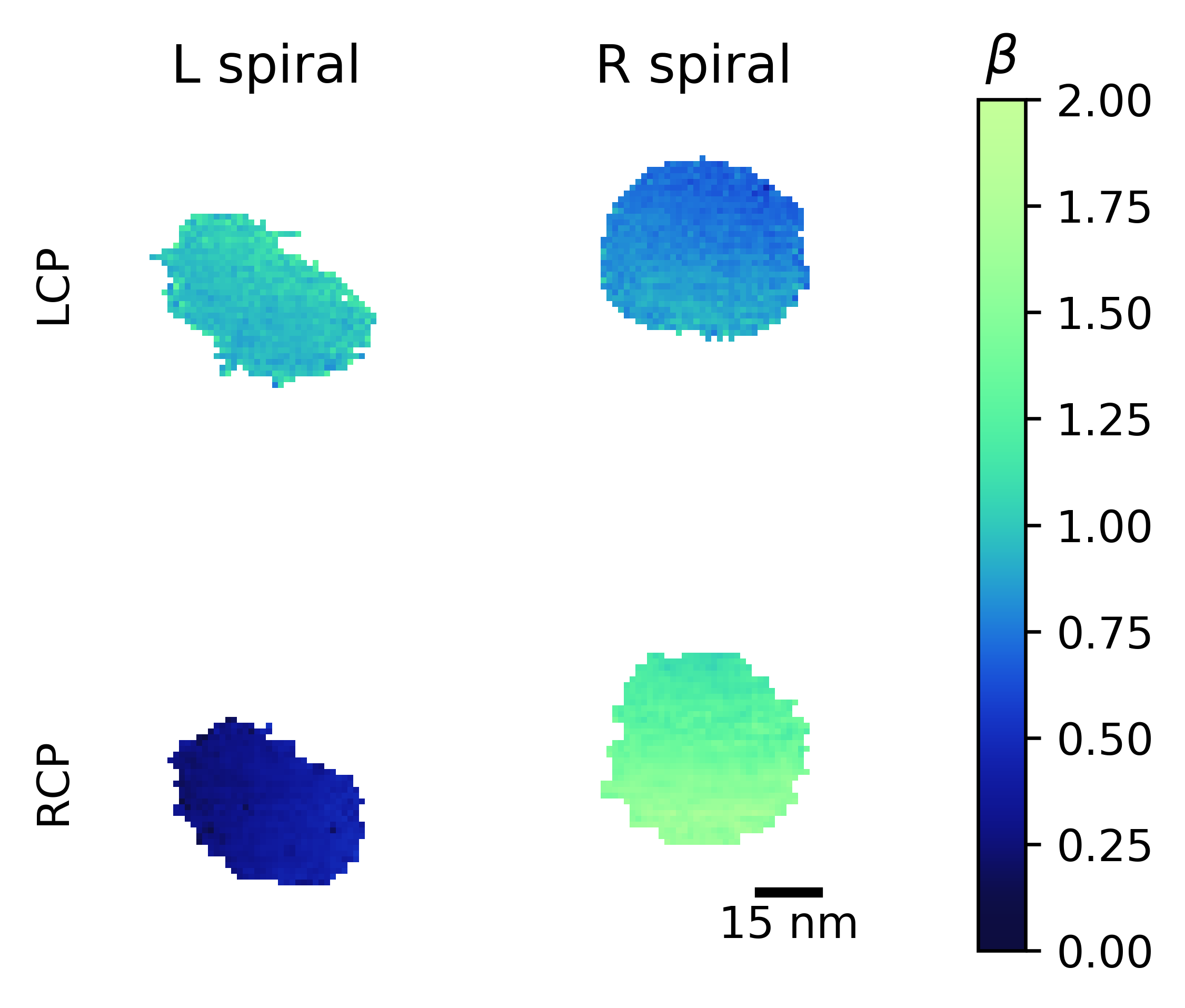}} \\
  \subfloat[]{\label{subfig:forest:dich}
  \includegraphics[height=0.8275in]{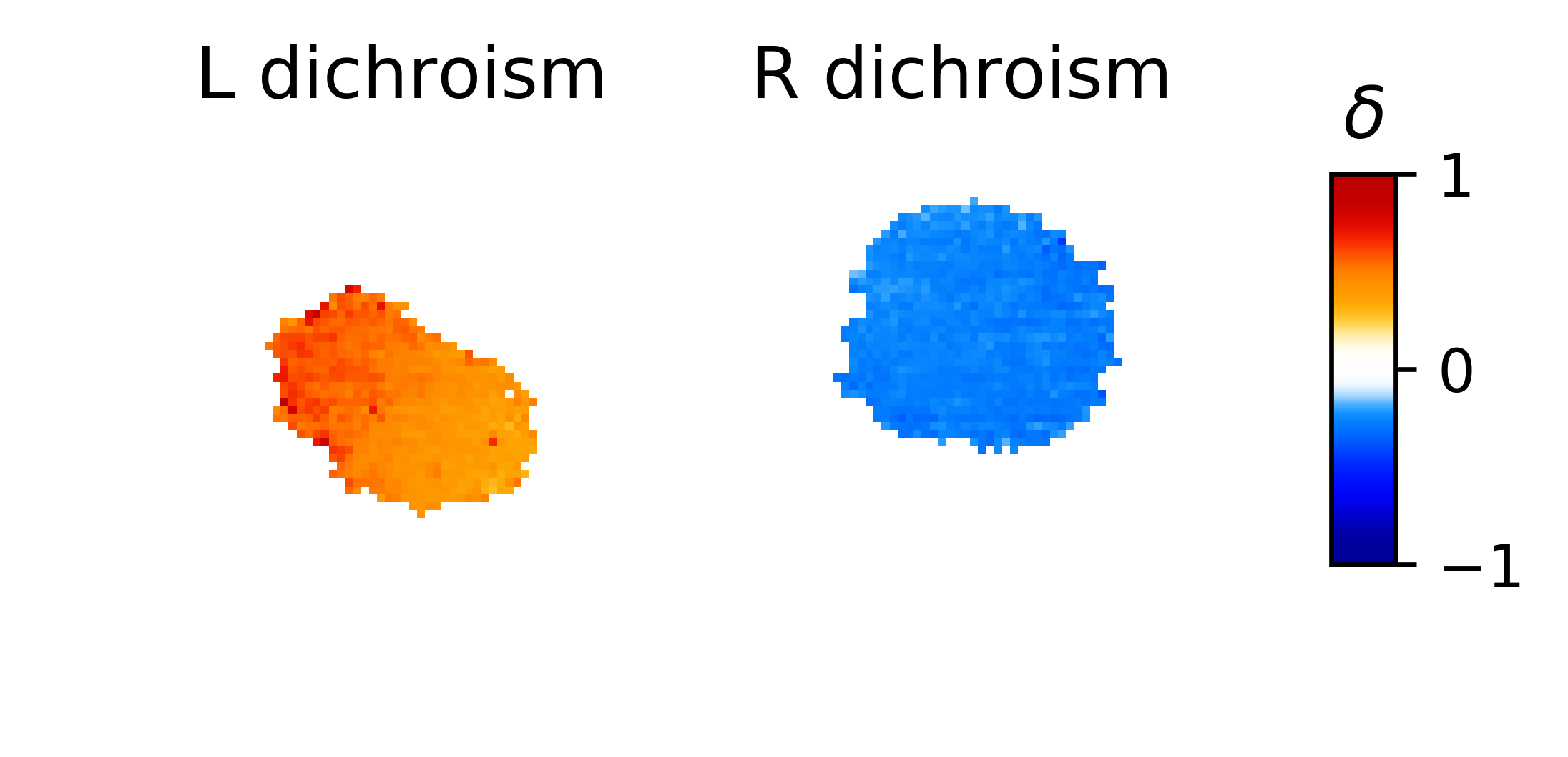}}
  \end{minipage}
  \caption{(a),(d) SEM images of structures with $\unit[400]{\textrm{nm}}$ pitch(structure in (a) same as shown in main text). (b),(e) Coupling constant $|\beta_{\sigma}(\mathbf{r}_{\perp})|$ measured with circular polarization for those structures. (c),(f) Relative circular dichroism $\Delta(\mathbf{r}_{\perp})$ for those structures. \label{fig:400circ}}
\end{figure}

\begin{figure}
  \begin{minipage}{0.49\textwidth}
    \subfloat[]{\label{subfig:teal:sem} 
  \includegraphics[height=0.5in]{teal_0deg_cropped_brightened.png}} \\
  \subfloat[]{\label{subfig:teal:beta}
  \includegraphics[height=1.5in]{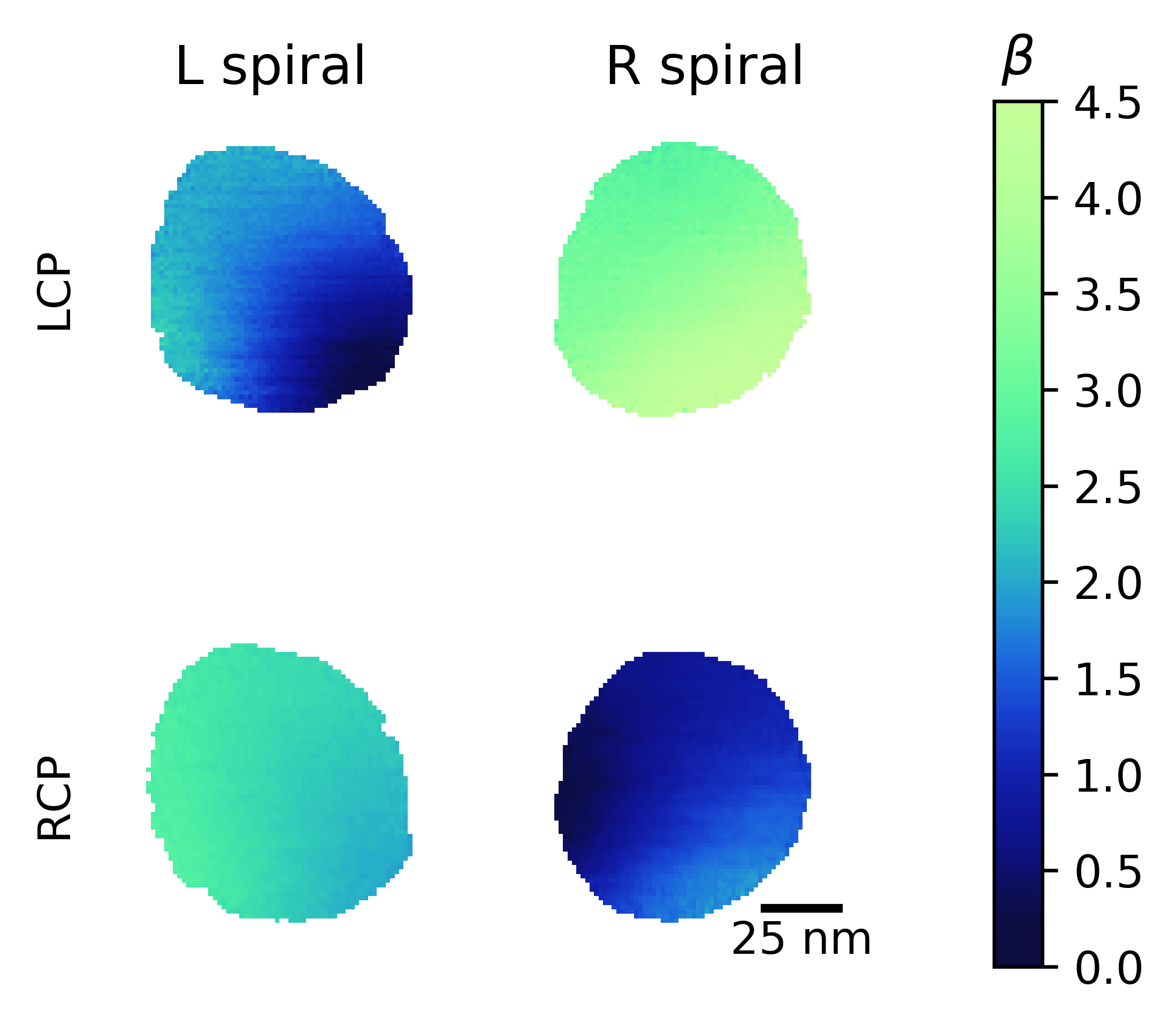}} \\
  \subfloat[]{\label{subfig:teal:dich}
  \includegraphics[height=0.8275in]{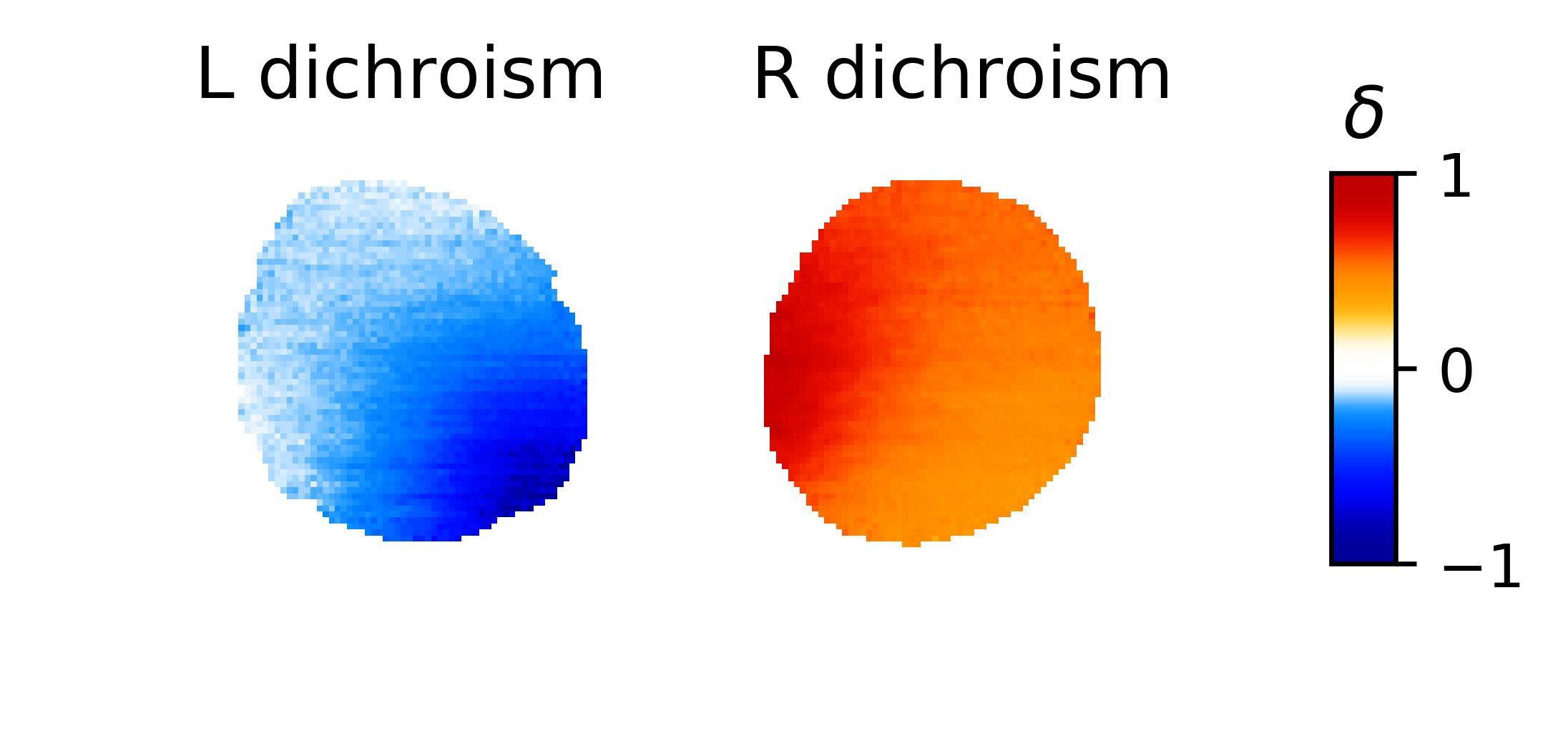}}
  \end{minipage}
  \begin{minipage}{0.49\textwidth}
    \subfloat[]{\label{subfig:orange:sem} 
  \includegraphics[height=0.5in]{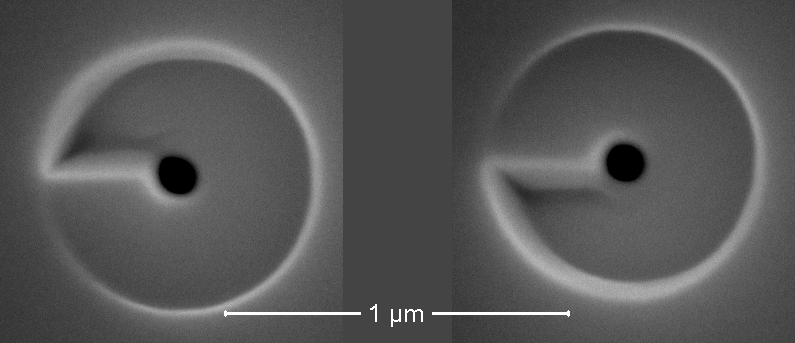}} \\
  \subfloat[]{\label{subfig:orange:beta}
  \includegraphics[height=1.5in]{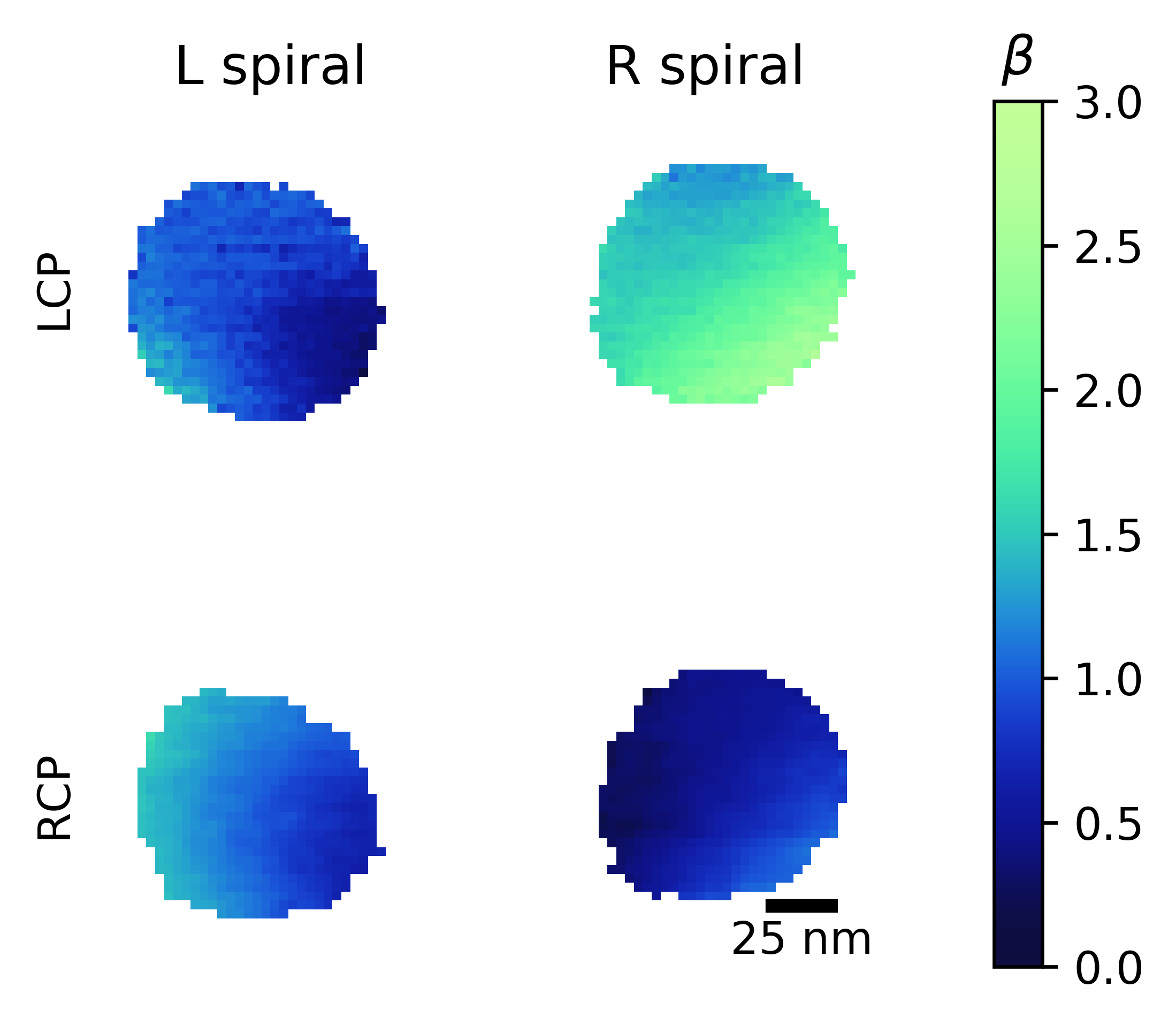}} \\
  \subfloat[]{\label{subfig:orange:dich}
  \includegraphics[height=0.8275in]{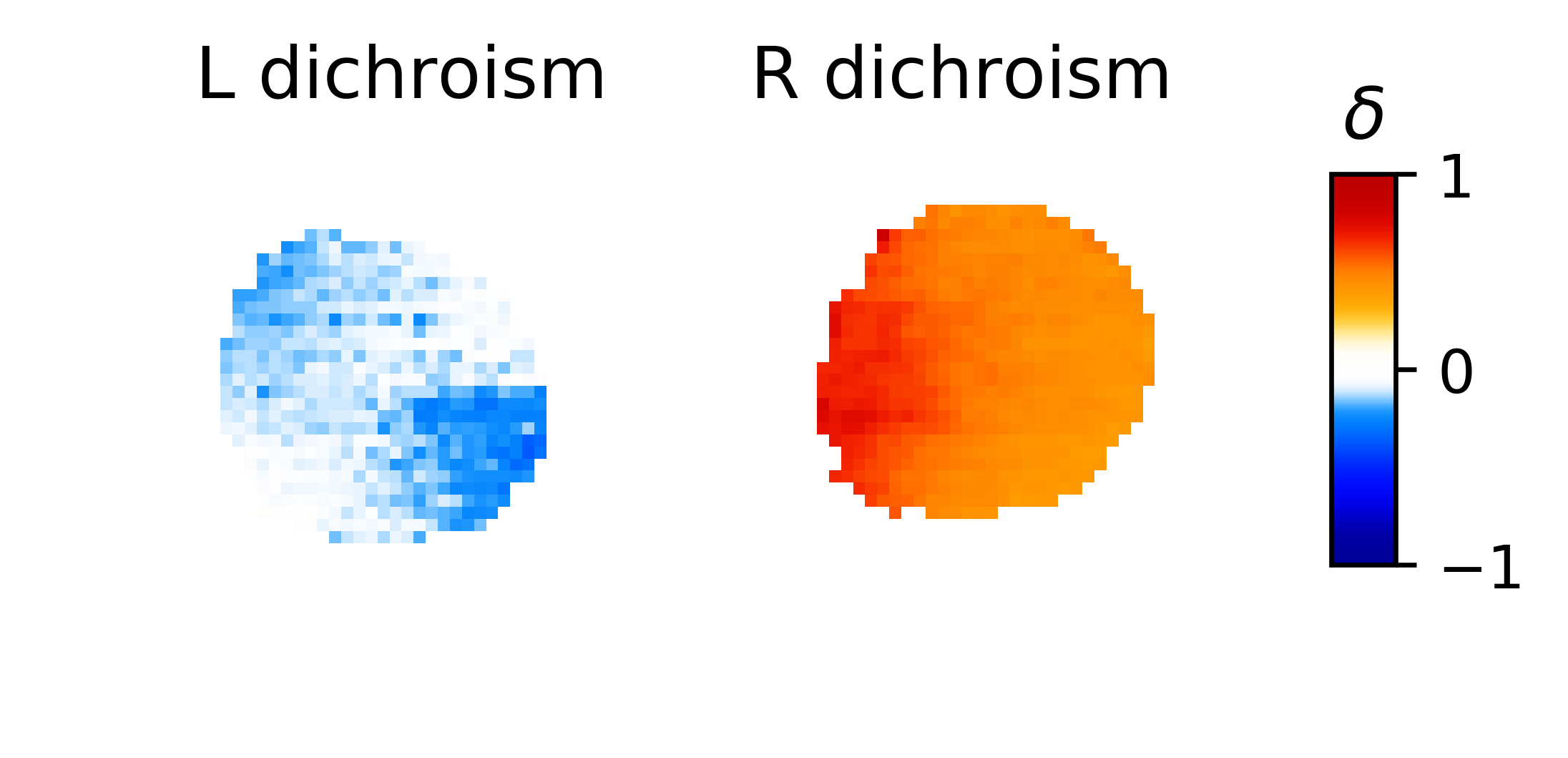}}
  \end{minipage} \\
  \begin{minipage}{0.49\textwidth}
    \subfloat[]{\label{subfig:blue:sem} 
  \includegraphics[height=0.5in]{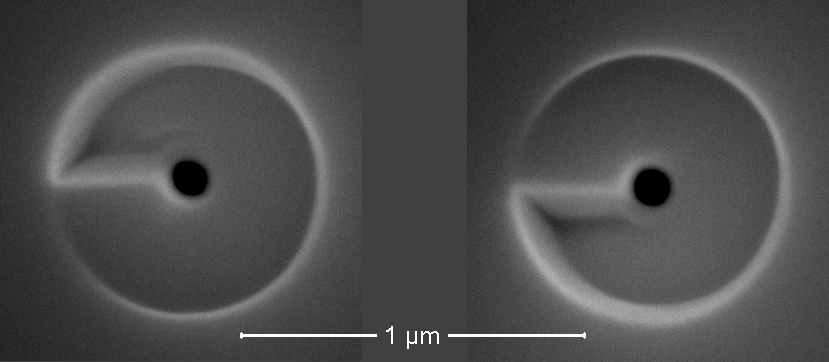}} \\
  \subfloat[]{\label{subfig:blue:beta}
  \includegraphics[height=1.5in]{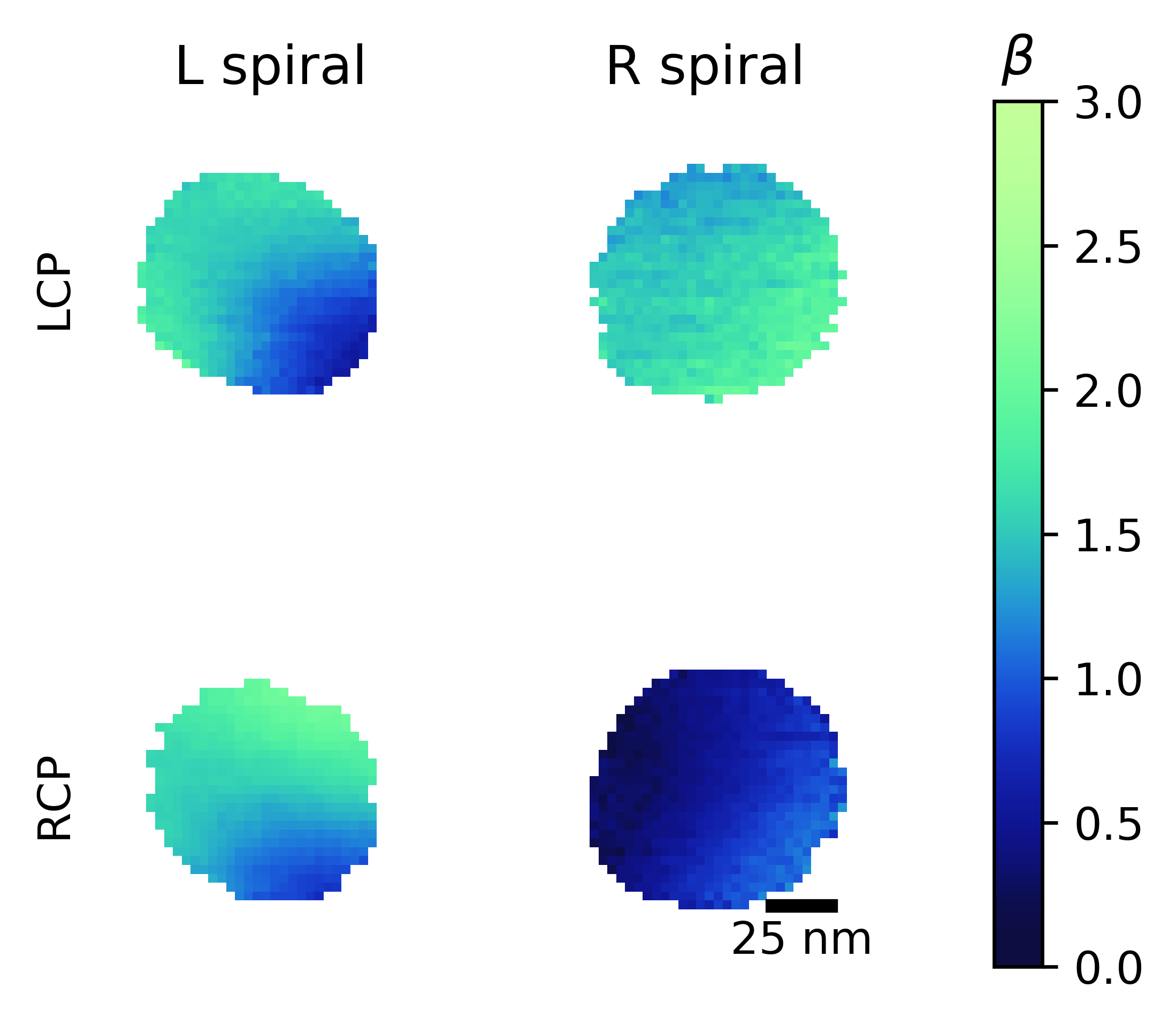}} \\
  \subfloat[]{\label{subfig:blue:dich}
  \includegraphics[height=0.8275in]{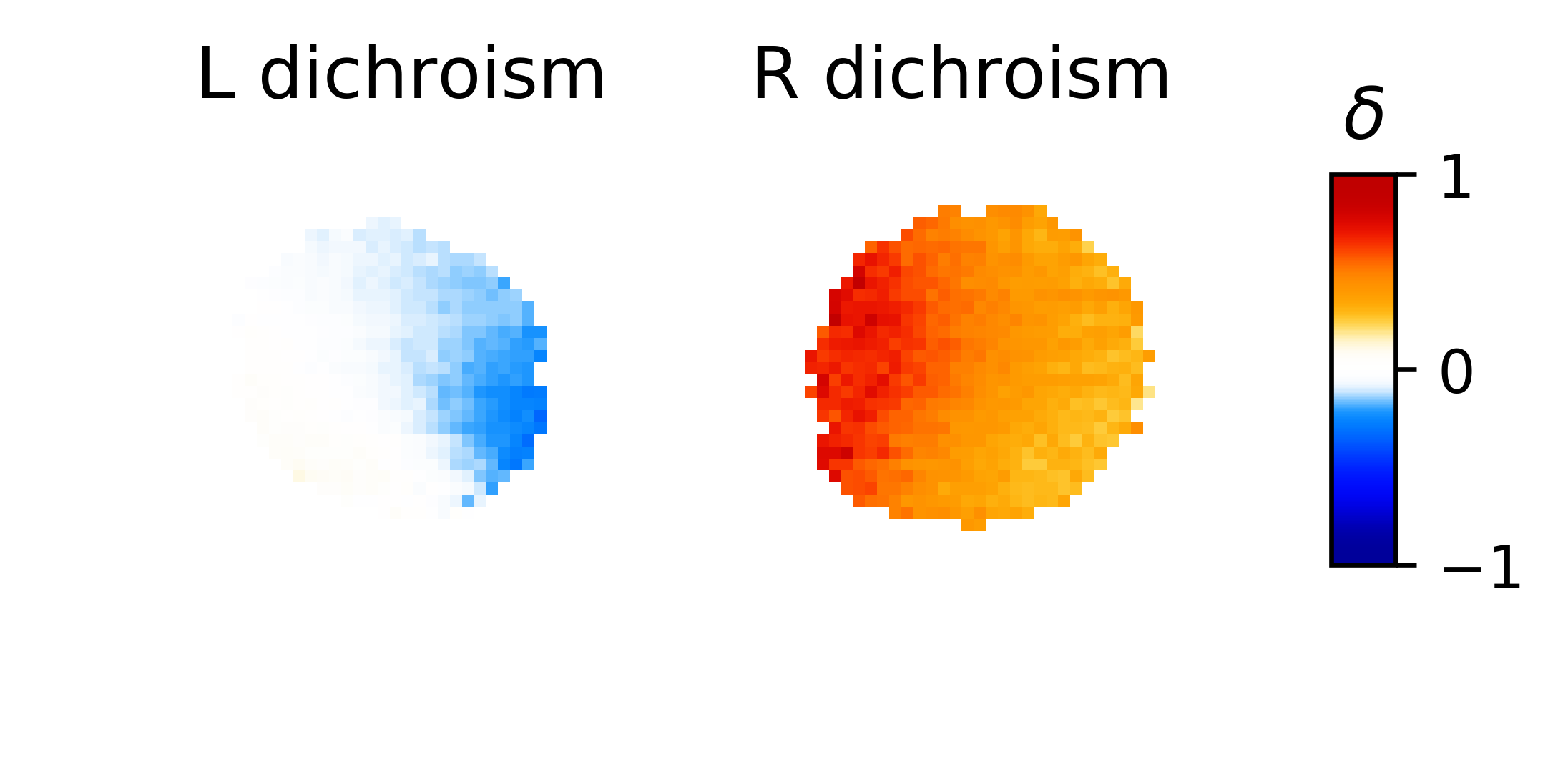}}
  \end{minipage}
  \begin{minipage}{0.49\textwidth}
    \subfloat[]{\label{subfig:yellow:sem} 
  \includegraphics[height=0.5in]{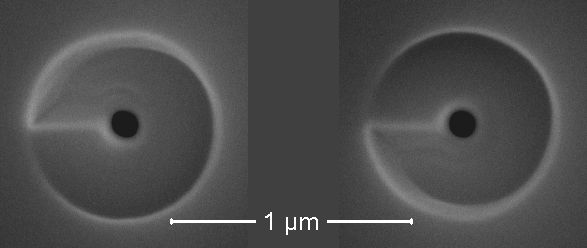}} \\
  \subfloat[]{\label{subfig:yellow:beta}
  \includegraphics[height=1.5in]{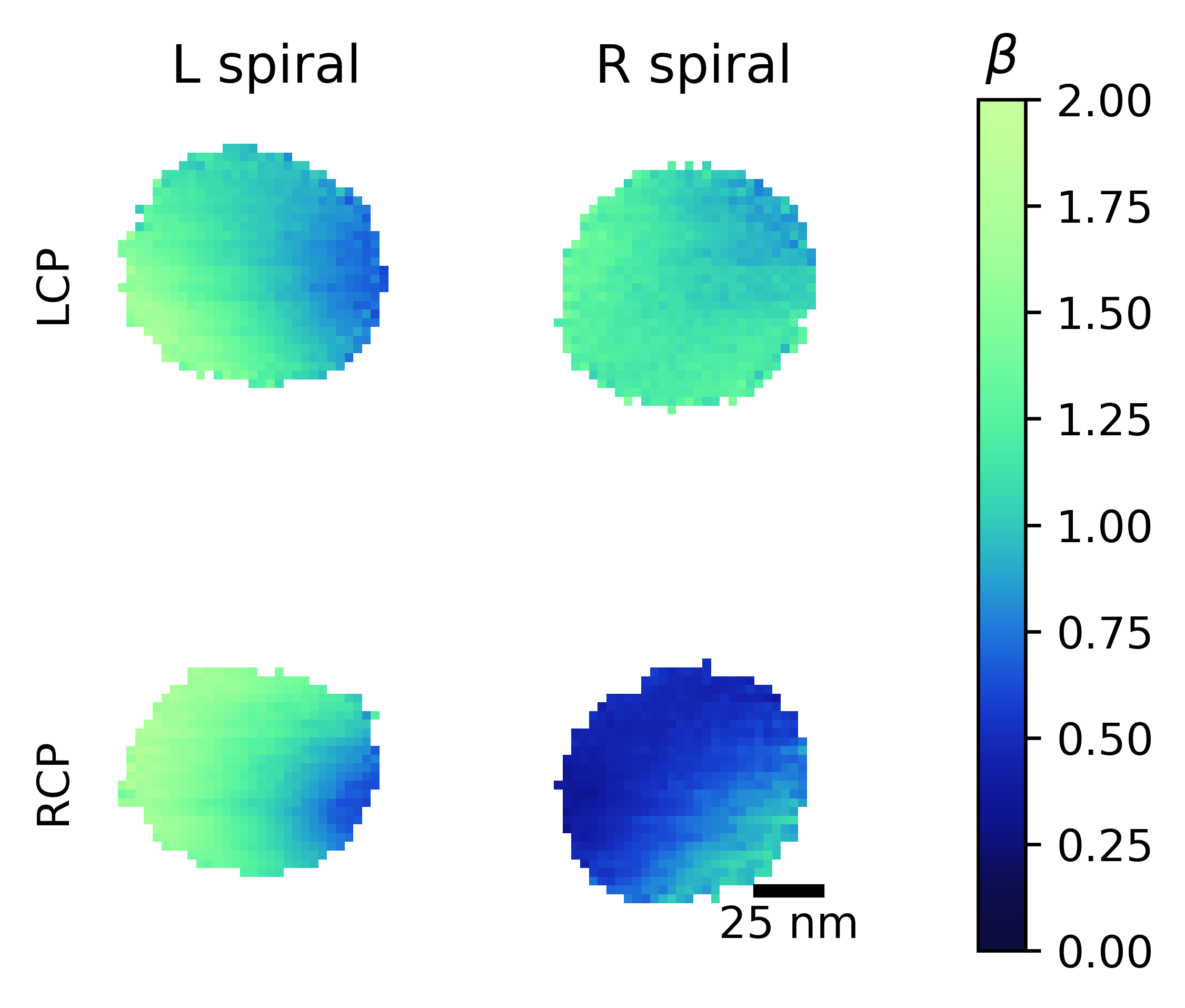}} \\
  \subfloat[]{\label{subfig:yellow:dich}
  \includegraphics[height=0.8275in]{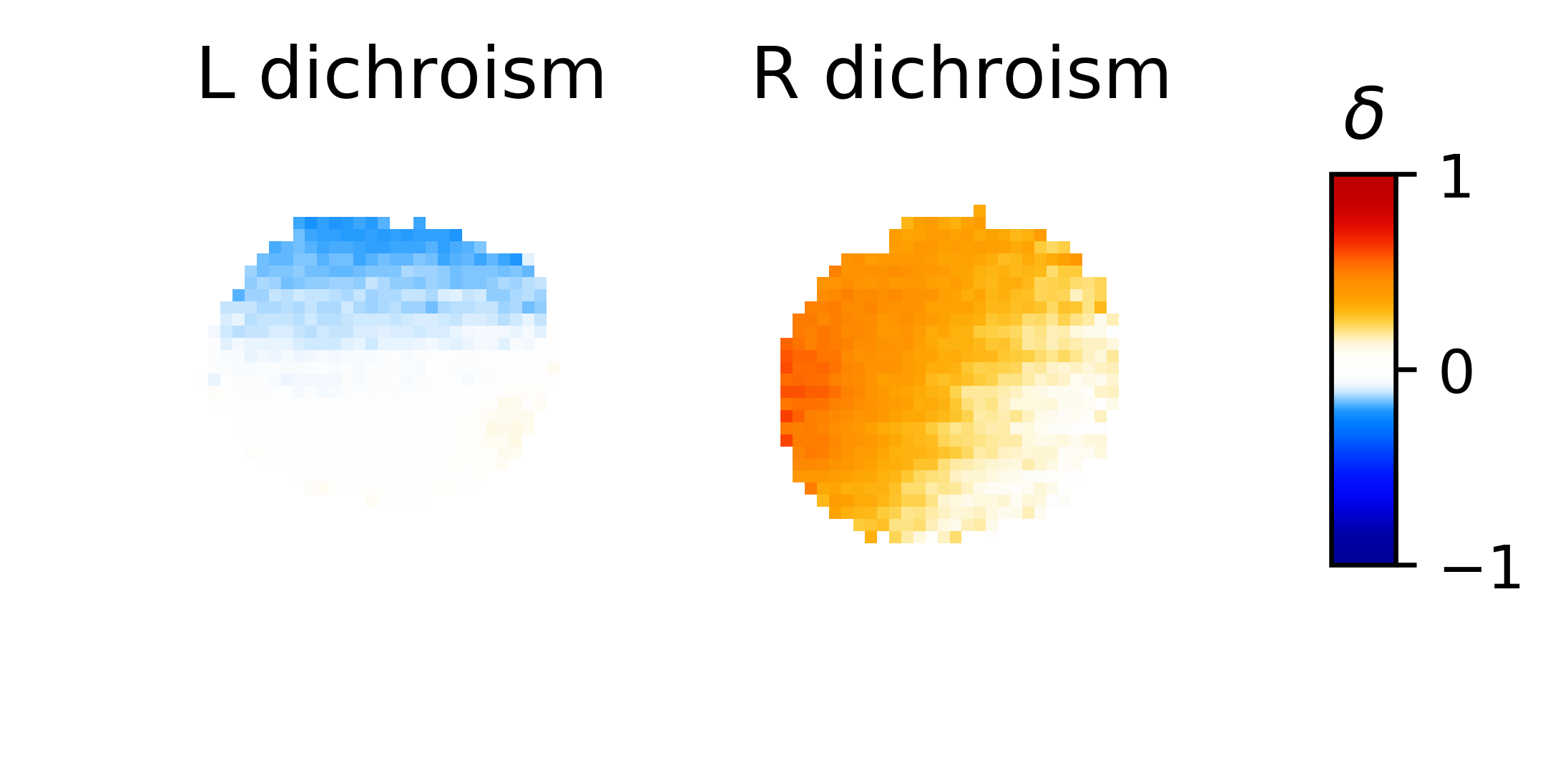}}
  \end{minipage}
  \caption{(a),(d),(g),(j) SEM images of structures with $\unit[800]{\textrm{nm}}$ pitch. (b),(e),(h),(k) Coupling constant $|\beta_{\sigma}(\mathbf{r}_{\perp})|$ measured with circular polarization for those structures. (c),(f),(i),(l) Relative circular dichroism $\Delta(\mathbf{r}_{\perp})$ for those structures. \label{fig:800circ}}
\end{figure}

To demonstrate the robustness of electron near-field circular dichroism, recorded circular and linear dichroism measurements on a wide array of FIB-milled spirals with both $\unit[400]{\textrm{nm}}$ and $\unit[800]{\textrm{nm}}$ pitch and varying through-hole diameters. The data consistently show positive circular dichroism for left-handed spirals with $\unit[400]{\textrm{nm}}$ pitch and negative circular dichroism for right-handed spirals (Fig. \ref{fig:400circ}). For $\unit[800]{\textrm{nm}}$ pitch spirals, we see positive dichroism for right-handed spirals and negative dichroism for left-handed spirals (Fig. \ref{fig:800circ}).  

We define linear dichroism as we did circular dichroism in Equation 2 in the main text:
\begin{equation}
  \Delta_{\leftrightarrow}(\mathbf{r}_{\perp}) = \frac{|\beta_{V}| - |\beta_{H}|}{|\beta_{V}|+|\beta{H}|},
\end{equation}
where $H$ signifies illumination with horizontally-polarized light (where horizontal here means the horizontal direction in all images we show), and $V$ signifies illumination with vertically-polarized light. The spatial average of linear dichroism $\Delta_{\leftrightarrow}$ on structures with cylindrical symmetry must be zero, just as the spatial average of circular dichroism $\Delta$ on non-chiral structures must also be zero.

\begin{figure}
  \begin{minipage}{0.4\textwidth}
  \subfloat[]{\label{subfig:red:lin_beta}
  \includegraphics[height=1.5in]{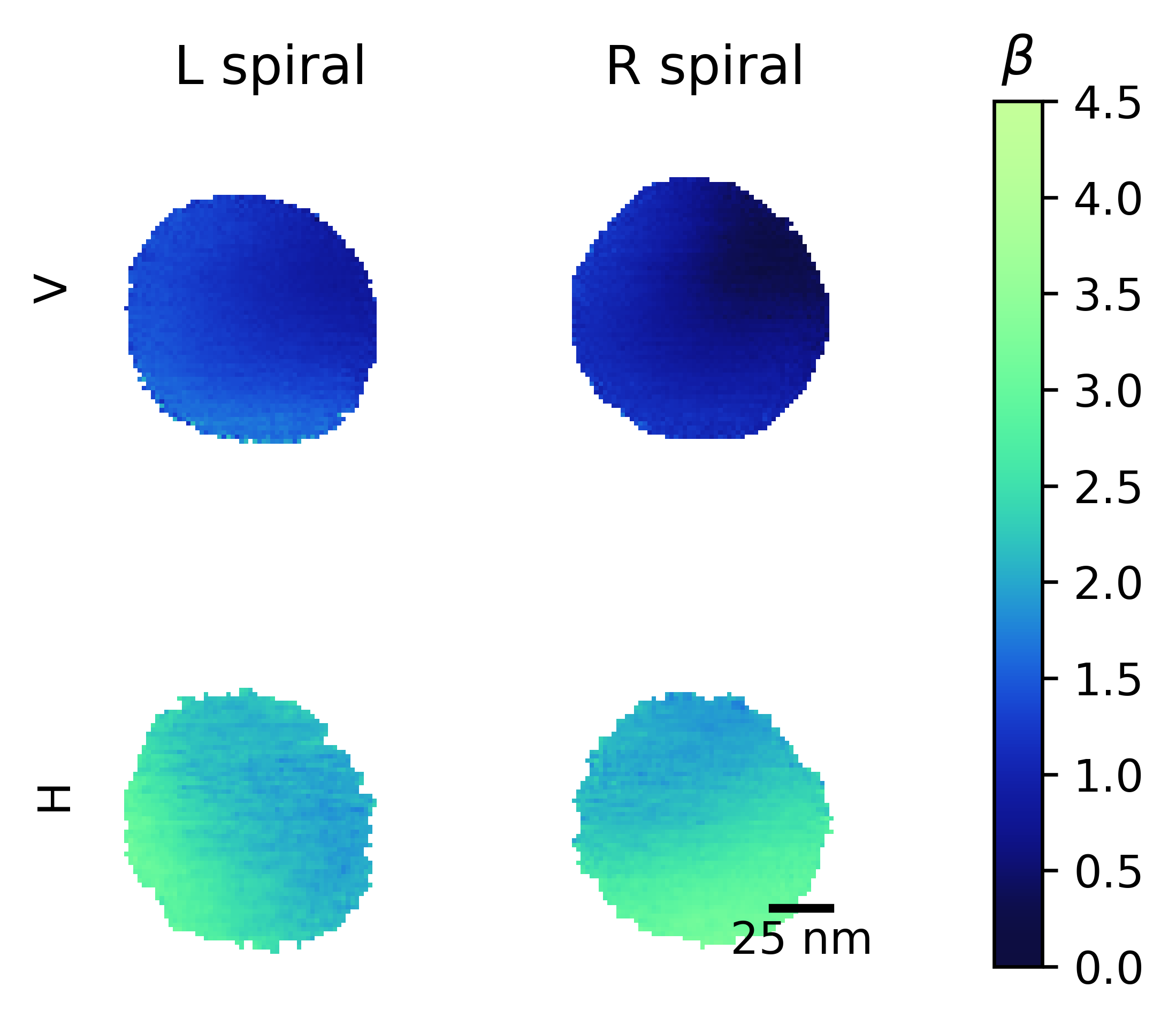}} \\
  \subfloat[]{\label{subfig:red:lin_dich}
  \includegraphics[height=0.8275in]{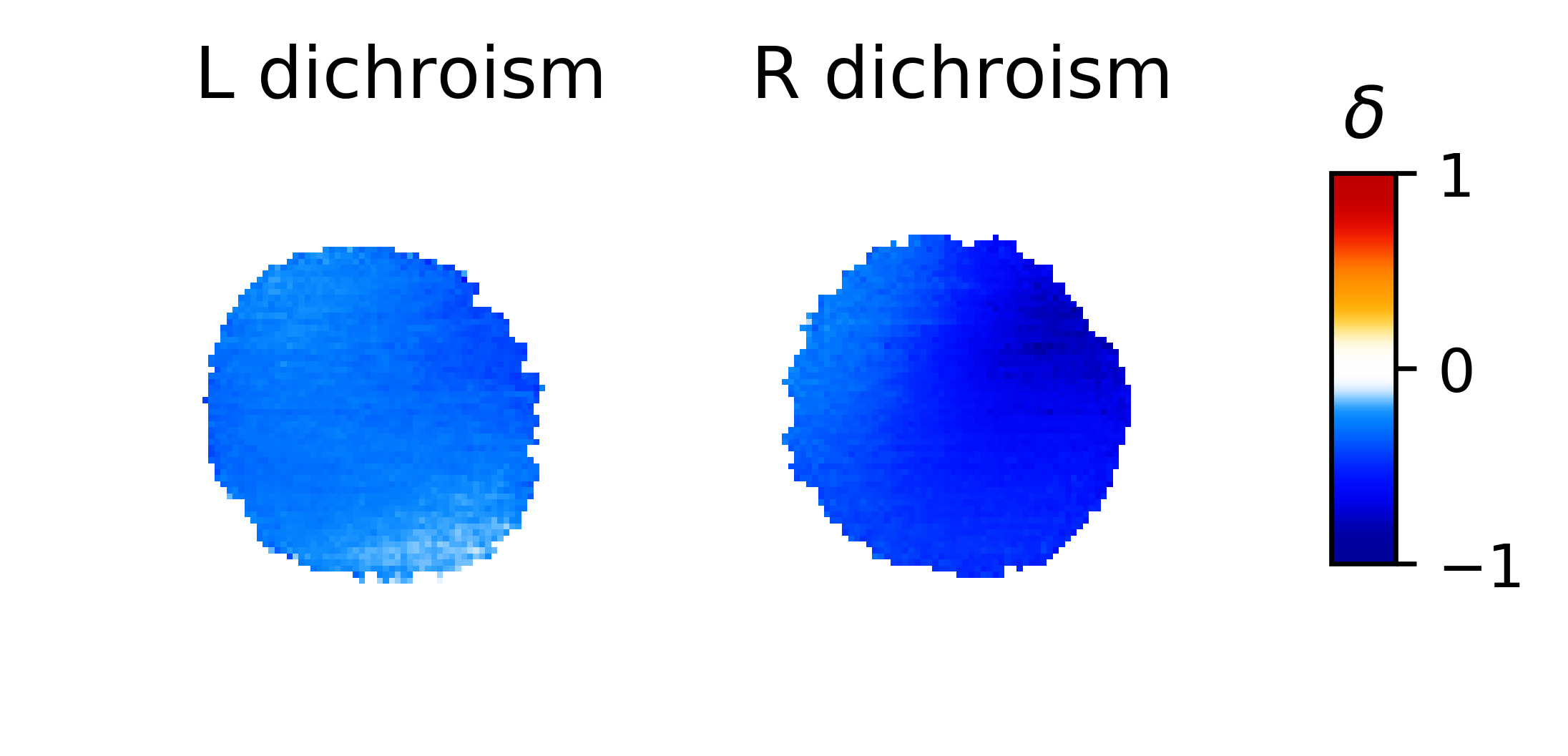}}
  \end{minipage}
  \begin{minipage}{0.4\textwidth}
  \subfloat[]{\label{subfig:teal:lin_beta}
  \includegraphics[height=1.5in]{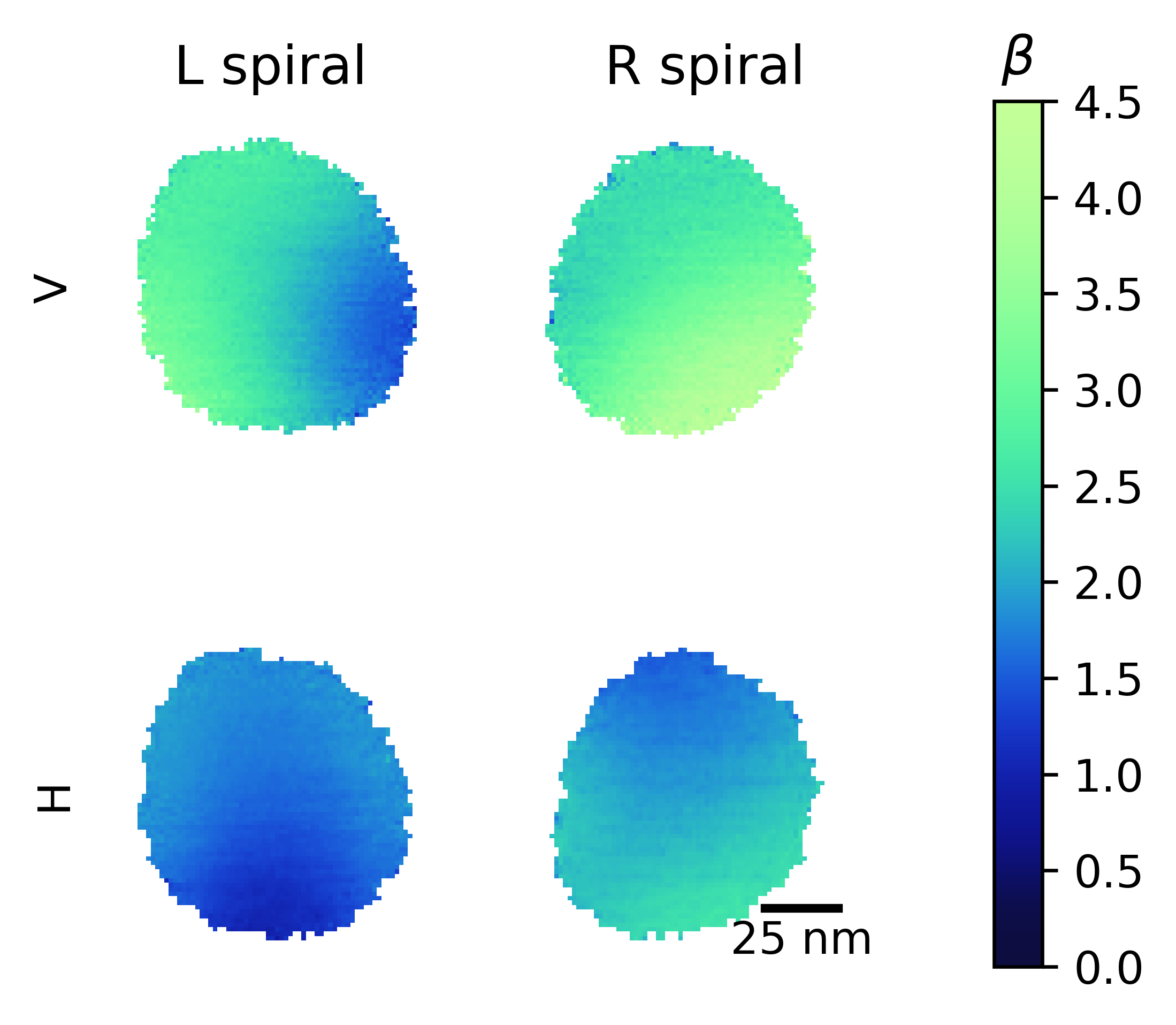}} \\
  \subfloat[]{\label{subfig:teal:lin_dich}
  \includegraphics[height=0.8275in]{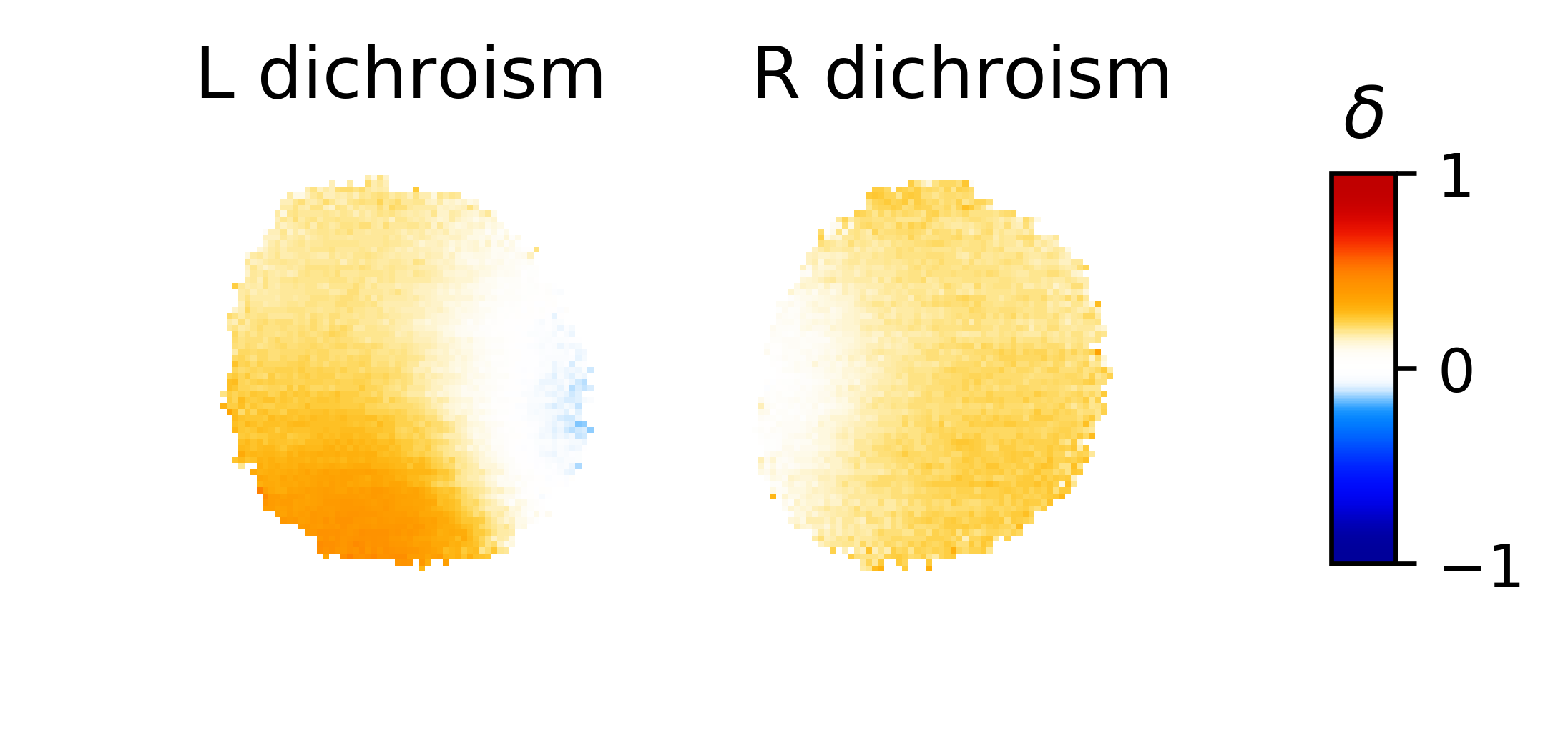}}
  \end{minipage}
  \caption{Coupling constant $\beta_{\sigma}(\mathbf{r}_{\perp})$ measured with linear polarization for structures with (a) $\unit[400]{\textrm{nm}}$ pitch (same structure as Fig. 1b in the main text) and (c) $\unit[800]{\textrm{nm}}$ pitch (same structure as Fig. \ref{subfig:teal:sem}). (b),(d) Relative linear dichroism $\Delta_{\leftrightarrow}(\mathbf{r}_{\perp})$ for those structures. The linear dichroism is independent of handedness of the structure, but depends on pitch in the same way as circular dichroism: all $\unit[400]{\textrm{nm}}$ pitch structures have opposite-sign dichroism of all $\unit[800]{\textrm{nm}}$ pitch structures. \label{fig:linear_both}}
\end{figure}

The pitch dependence we see for circular dichroism also appears in the linear dichroism measurements. As expected, we see in Fig. \ref{fig:linear_both} nearly identical linear dichroism on left- and right-handed spirals, but the sign of the linear dichroism flips when spiral pitch is increased to $\unit[800]{\textrm{nm}}$.

\section{Polarization calibration}

As the optical pump path includes three mirrors between the quarter- and half-wave plate we used to define the pump polarization and the specimen, we measured the Jones matrix of these three optical elements in order to produce circular polarization at the specimen. To do so, we recorded PINEM maps in a $\unit[0.6]{\mu\textrm{m}}$ by $\unit[0.6]{\mu\textrm{m}}$ area at the corner of a silicon nitride window on a silicon frame. The two nearly-othogonal edges respond most strongly to linear polarization perpendicular to the edge, and a small region in the corner is sensitive to the relative phase of the two orthogonal linear polarizations \cite{echternkamp_ramsey-type_2016}. We recorded spatial maps of the coupling constant for a total of 162 combinations of quarter-wave plate and half-wave plate angles evenly spaced between $-90$ and $90\degree$ and $-45$ and $45\degree$, respectively. We found the Jones matrix that minimized least-squares error between our predicted PINEM maps and measured maps for all waveplate points. We then used this reconstructed Jones matrix to identify waveplate angles that produce left- and right-circular polarization at the specimen.

\section{Coupling constant measurement \label{sect:coup_meas}}

In the presence of inelastic electron-light interaction with coupling constant $\beta$, the electron's probability for gain or loss of $m$ quanta of photon energy and momentum is \cite{garcia_de_abajo_multiphoton_2010,park_photon-induced_2010,feist_quantum_2015}
\begin{equation}
  P_m = |J_m(2|\beta|)|^2,
\end{equation}
where $J_m$ is the $m$th-order Bessel function of the first kind.
When the initial electron energy distribution $n(E)$ is dominated by incoherent fluctuations, the probability distribution for the energy of an electron is
\begin{equation} \label{eq:p_E}
  p(E) = |\psi_f(E)|^2 \otimes n(E)  = \left(\sum_m |J_m(2|\beta|)|^2 \delta(E-m \Eph)\right) \otimes n(E), 
\end{equation}
where $\psi_f(E)$ is the coherent electron wavefunction after inelastic electron-light interaction, and $\otimes$ denotes a convolution in energy. The measured energy spectrum of an average of $N_e$ electrons per spectrum in the presence of spatial or temporal averaging of the coupling constant is
\begin{equation} \label{eq:model}
  I_{\Eph,E_0,\beta,\Delta \beta,N_e}(E) = N_e \sum_{m=-\infty}^{\infty}  \left[f_{\Delta \beta}(\beta) \otimes J_m(2|\beta|)^2\right] n(E-m \Eph-E_0),
\end{equation}
 where $n(E-E_0)$ is the electron energy distribution measured in vacuum, also known as the zero-loss peak, $E_0$ is the apparent energy offset of the spectrum on the spectrometer camera, $\otimes$ here denotes a convolution in coupling constant, and the distribution 
\begin{equation}
  f_{\Delta \beta}(\beta) = \frac{1}{\sqrt{2\pi} \Delta \beta} \ee^{-\frac{1}{2}\left(\frac{\beta}{\Delta \beta}\right)^2}
\end{equation}
accounts for spatial or temporal averaging of the coupling constant with a width $\Delta \beta$. We cut off the convolution at $\beta=0$ since the energy distribution we measure depends only on $|\beta|$.

We measured the coupling constant at each spatial position by least-squares minimization of the difference between the measured electron energy spectra and this model. 

\section{Finite element method simulations}
We used the \textit{COMSOL Wave Optics Module} \cite{comsoluserguide} to find a numerical solution for the optical near-field of the spiral structures milled into gold. In particular, we used the scattered field formulation in the frequency domain and divided the simulation into two parts as explained in reference \citep{comsolguidescatterer}. This method has, for instance, been applied by Piazza et al. for the determination of the near fields of silver nanowires \citep{piazza2015comsol}.

The first step is to solve the wave equation in the rectangular simulation volume, including a gold block with a cylindrical hole of radius $\unit[400]{nm}$ and depth $\unit[800]{nm}$. This step uses Dirichlet boundary conditions, known as Port boundary conditions in COMSOL, at the top and bottom of the volume and Floquet periodic boundary conditions for the sides. The second step employs the solution of the first step as a background field to calculate the scattered field from interaction with the spiral with perfectly matched layer boundary conditions. We use a mesh size in both steps such that the maximum distance between two vertices is always less than a sixth of the excitation wavelength in air and a tenth of the wavelength in the gold structure.

We numerically calculate the coupling constant $\beta$ based on the calculated $z$-component of the electric field \footnote{A factor of 2 difference can appear in definitions of $\beta$ \cite{garcia_de_abajo_multiphoton_2010,feist_quantum_2015} depending on whether the electric field is treated as real or complex.}, for electrons propagating in the $\vec{\hat{z}}$ direction according to Equation 1 in the main text,
\begin{equation} \label{eq:beta}
  \beta_{\sigma,\wph}(\mathbf{r}_{\perp}) = \frac{e}{\hbar \wph} \int \mathrm{d}z \ E_z(\mathbf{r},\sigma,\wph) \ee^{-i\wph z / \ve},
\end{equation}
where the frequency of the optical field is $\wph = \frac{2 \pi c}{\lambda}$, with $\lambda = \unit[800]{nm}$, and the electron velocity is $\ve = c \sqrt{1- \frac{1}{\left(1-\frac{E}{\me c^2}\right)}}$ with electron energy  $E = \unit[200]{keV}$, electron mass $\me$ and speed of light $c$. We scaled the intensity to match the magnitude of $\beta$ measured in experiment, as we measure incident power outside the electron microscope and the spot profile at the specimen is not rigorously characterized. 

\section{Models for dichroism maps}

In this section, we develop two models to explain dichroism in the coupling constant. First, we describe the coupling constant produced by oscillating electric and magnetic dipoles. Then, we construct a model for the specimen we considered in this paper by treating the inner edge of the hole in the spiral as a 1D magnetic polarizability. 

\subsection{Electron coupling to an electric dipole}

Following a similar procedure as described elsewhere,\cite{giulio_probing_2019} we write the electric field produced at a position $\rb$ by an electric dipole $\pb$ located at $\rb_0$ and oscillating with frequency $\wph$ using Einstein summation convention as
\begin{align}
  E_{\ell}(\rb) = \left(k^2 \delta_{\ell m} +\partial_{\ell} \partial_m \right) p_m \frac{\ee^{\left(i k |\rb-\rb_0|\right)}}{|\rb-\rb_0|} ,
\end{align}
where $k=\wph/c$. Using this field in Eq.\ \eqref{eq:beta}, we find
\begin{align}
  \beta_{\rb_0,\pb}(\rb_\perp) &= \frac{2e}{\hbar\wph} \left(k^2 p_z + \pb\cdot\nabla_{\rb_0} \partial_{z_0}\right) K_0{(qd_\perp/\gamma_e}) \ee^{-iqz_0} \nonumber\\
  &= \frac{2e}{\hbar\wph} \left[(k^2-q^2) p_z K_0{(qd_\perp/\gamma_e)} +\frac{\pb\cdot\db_\perp}{d_\perp} \frac{iq^2}{\gamma_e} K_1{(qd_\perp/\gamma_e)} \right] \ee^{-i q z_0} \nonumber\\
  &=\frac{2e \wph}{\hbar \ve^2 \gamma_e} \left[\frac{i\pb\cdot\db_\perp}{d_\perp}  K_1{\left(\frac{\wph d_\perp}{\ve\gamma_e}\right)}- \frac{p_z}{\gamma_e} K_0{\left(\frac{\wph d_\perp}{\ve\gamma_e}\right)}\right]\ee^{-i\wph z_0/\ve}, 
\end{align}
where $q=\wph/\ve$, $\gamma_e = 1/\sqrt{1-\ve^2/c^2}$, and $\db_{\perp}=\rb_{0 \perp}-\rb_{\perp}$.

\subsection{Electron coupling to a magnetic dipole}

The electric field procuced by a magnetic dipole $\mb$ is
\begin{align}
  \Eb(\rb) = ik\,\mb\times\nabla_{\rb_0} \frac{\ee^{ik|\rb-\rb_0|}}{|\rb-\rb_0|},
\end{align} 
which, upon insertion in Eq.\ \eqref{eq:beta}, produces a coupling coefficient
\begin{align}
  \beta_{\rb_0,\mb}(\rb_{\perp}) &= \frac{2i e}{\hbar c}\, \zz\cdot \left(\mb\times\nabla_{\rb_0}  \right) K_0{(qd_\perp/\gamma_e)} \ee^{-iqz_0} \nonumber\\
  &= -\frac{2i e}{\hbar c} \zz\cdot \left(\mb\times\db_{\perp}\right)\frac{q}{\gamma_e d_{\perp}} K_1{(qd_\perp/\gamma_e)} \ee^{-iqz_0}  \nonumber\\
  &= -\frac{2i \wph e}{\hbar c \ve \gamma_e d_{\perp}}  K_1{\left(\frac{\wph d_\perp}{\ve\gamma_e}\right)}\;\zz\cdot\left(\mb\times\db_{\perp}\right) \ee^{-i\wph z_0/\ve}. \label{eq:magnetic_dipole}
\end{align}

\subsection{Set of magnetic and electric dipoles}

For a set of electric and magnetic dipoles $\pb_i$ and $\mb_j$ at positions $\mathbf{r}_{0i}$ and $\mathbf{r}_{0j}$,
\begin{align} 
  \beta_{\pb,\mb}(\rb_{\perp}) &= \frac{2 e \wph}{\hbar \ve^2 \gamma_e} \sum_i \left[ i \pb_i \cdot \db_{\perp i} K_1{\left(\frac{\wph d_{\perp i }}{\ve \gamma_e}\right)} - \frac{p_{zi}}{\gamma_e} K_0{\left(\frac{\wph d_{\perp i}}{\ve \gamma_e}\right)}\right] e^{-i\wph z_{0i} / \ve} \nonumber \\
  &\phantom{=}- \frac{2 i \wph e}{\hbar c \ve \gamma_e d_{\perp}} \sum_j K_1{\left(\frac{\wph d_{\perp j}}{\ve \gamma_e}\right)}\; \zz \cdot \left(\mb_j \times \db_{\perp j} \right)\ee^{-i \wph z_{0j}/\ve}. \label{eq:dipoles}
\end{align}

The coupling constant map for a small molecule with a set of discrete electronic transitions should be modeled well outside the molecule by Eq.\ \eqref{eq:dipoles}. We can see that, as is the case for far-field light \cite{power_circular_1974}, non-orthogonal electric and magnetic dipole transition moments with a $\pi/2$ phase difference will produce ENFCD.

\section{1D polarizable helix \label{sect:model}}

As the optical properties of the specimens we used are not dominated by a small number of electronic transitions, we sought to model the structure with a one-dimensional shape with magnetic and electric polarizability. At the $\unit[800]{\textrm{nm}}$ optical wavelength we used, the optical response of gold can be treated as a good conductor with a small skin depth compared with the size of the hole in the spiral. We do not expect strong resonant plasmons in this open structure, and the hole size is also below the threshold needed to sustain a guided mode \cite{bethe_theory_1944}. We therefore could apply past work on holes drilled in perfect conductors, which are known to produce an optical response consisting of an in-plane magnetic dipole and an out-of-plane electric dipole \cite{garcia_de_abajo_colloquium_2007}. 

We treated the inner edge of the hole in our gold spirals as a set of magnetic dipoles \footnote{As the hole diameter in our spirals is much smaller than the wavelength, the optical behavior is dominated by a quasistatic magnetic response.} pointing parallel to the edge. We therefore defined a magnetic polarizability per unit length $M$ along the inner edge of the hole in the spiral. The magnetic dipole induced by an external optical magnetic field $\Hb$ acting on an element of edge length $\text{d}l$ is $\text{d}\mb=M \text{d}l(\Hb\cdot\hat{\bf t})\,\hat{\bf t}$, where $\hat{\bf t}$ is the unit vector along the edge at the position under consideration. For simplicity, we fit the value of $M$ so that for a planar hole of radius $a$ the total induced dipole (i.e., integrated along along the circular edge) matches the result obtained for a deep hole drilled in a perfect conductor \cite{garcia_de_abajo_colloquium_2007}, which leads to $M=0.0475\,a^2$.

\begin{figure}
  \begin{minipage}{0.19\textwidth}
    \subfloat[]{\label{subfig:model:LCP_1} 
  \includegraphics[height=1in]{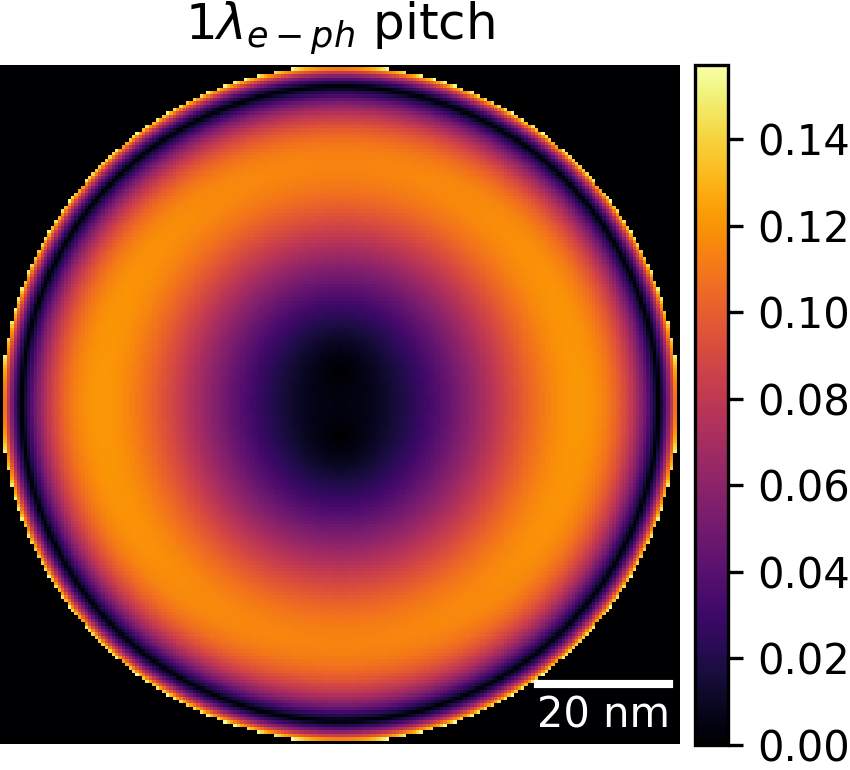}} \\
  \subfloat[]{\label{subfig:model:RCP_1}
  \includegraphics[height=1in]{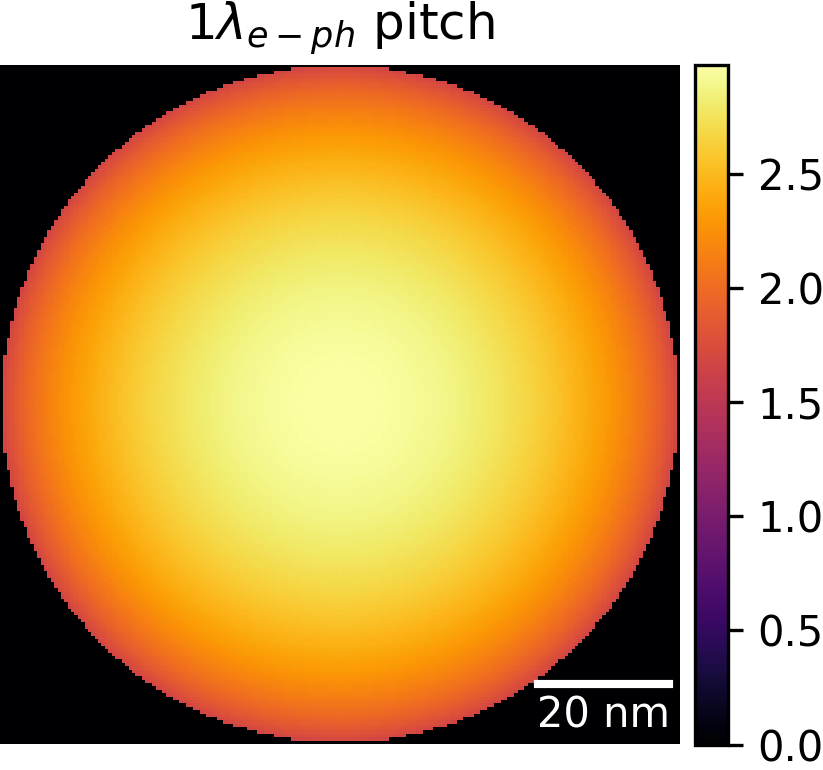}} \\
  \end{minipage}
  \begin{minipage}{0.19\textwidth}
    \subfloat[]{\label{subfig:model:LCP_2} 
  \includegraphics[height=1in]{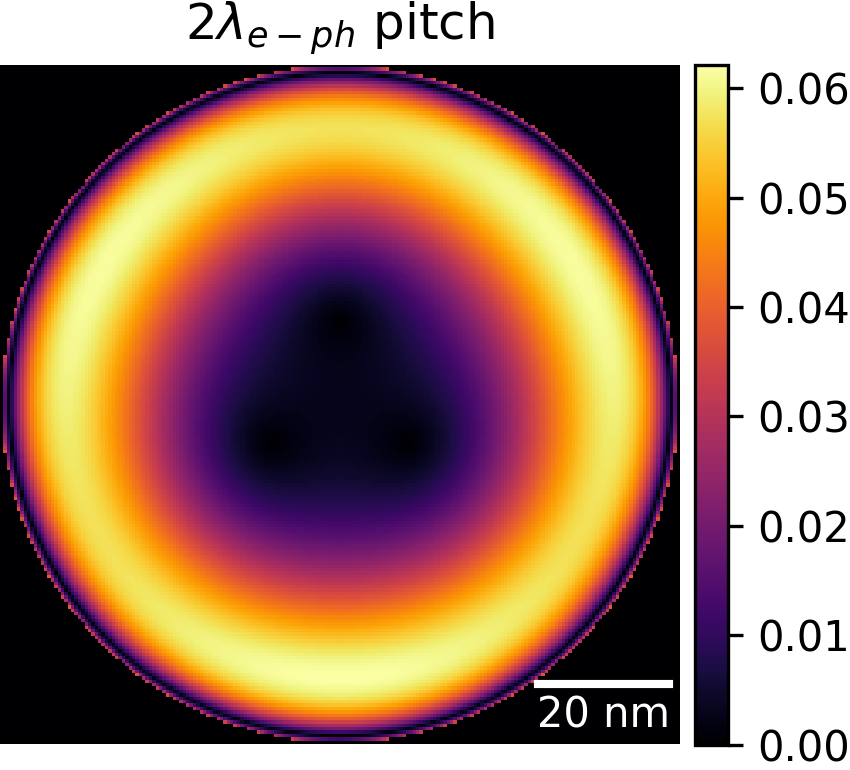}} \\
  \subfloat[]{\label{subfig:model:RCP_2}
  \includegraphics[height=1in]{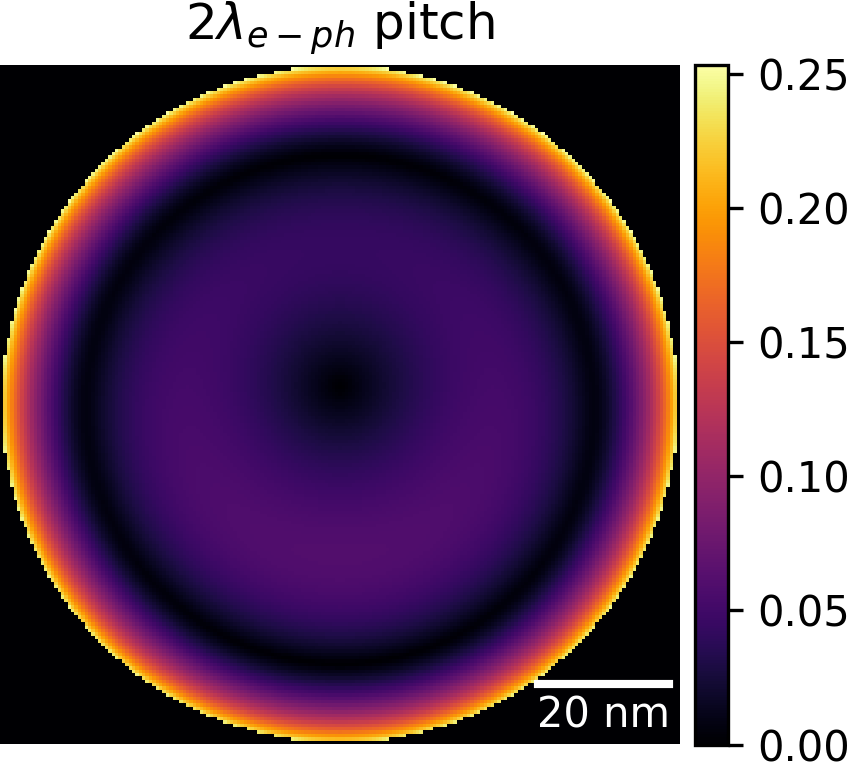}} \\
  \end{minipage}
  \begin{minipage}{0.19\textwidth}
    \subfloat[]{\label{subfig:model:LCP_3} 
  \includegraphics[height=1in]{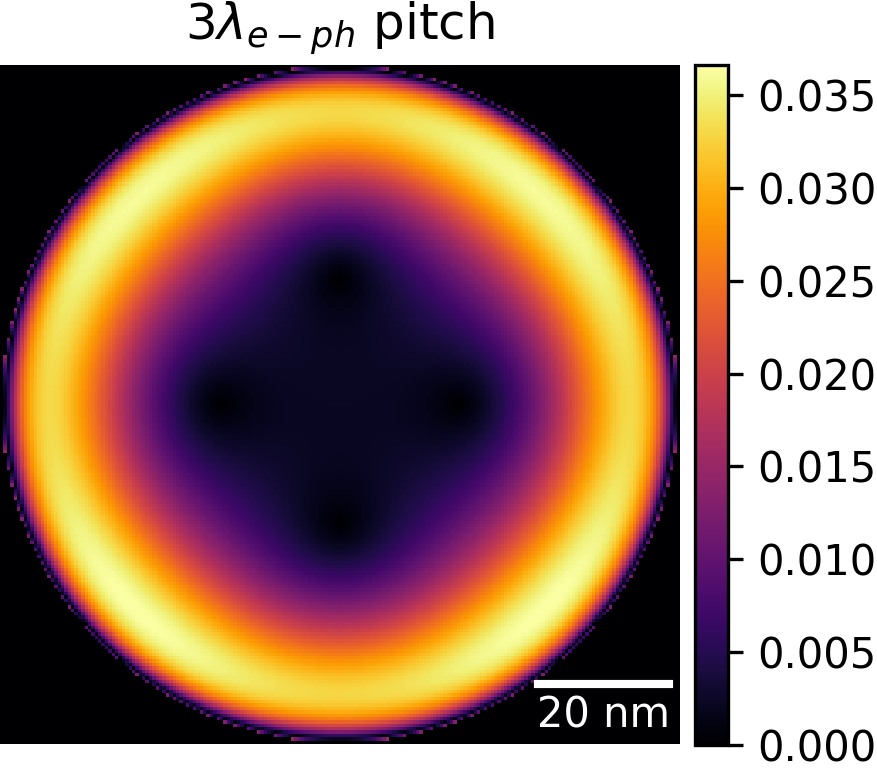}} \\
  \subfloat[]{\label{subfig:model:RCP_3}
  \includegraphics[height=1in]{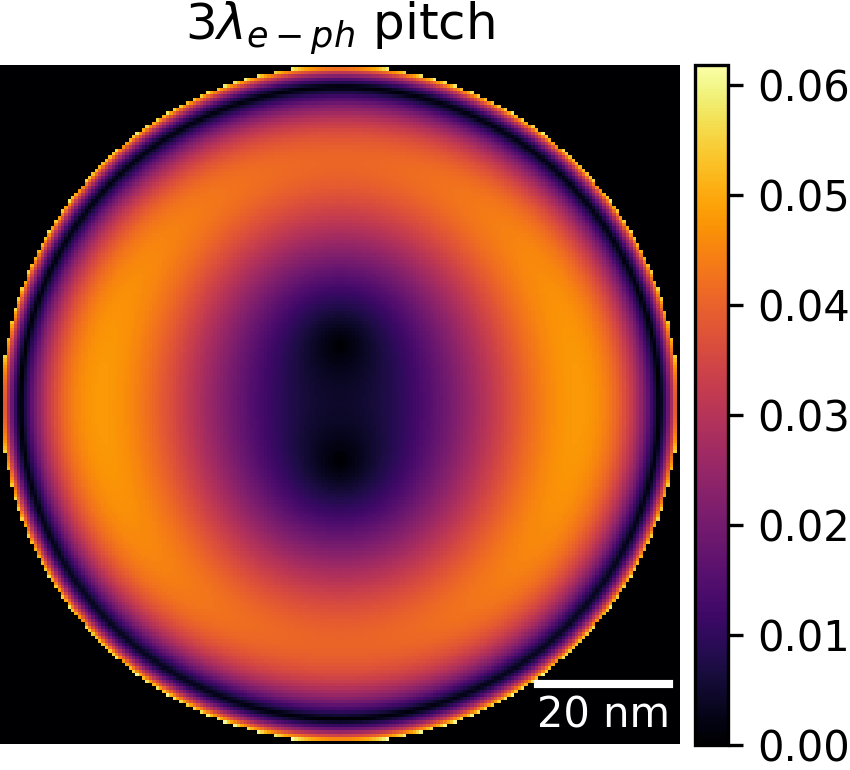}} \\
  \end{minipage}
  \begin{minipage}{0.19\textwidth}
    \subfloat[]{\label{subfig:model:LCP_4} 
  \includegraphics[height=1in]{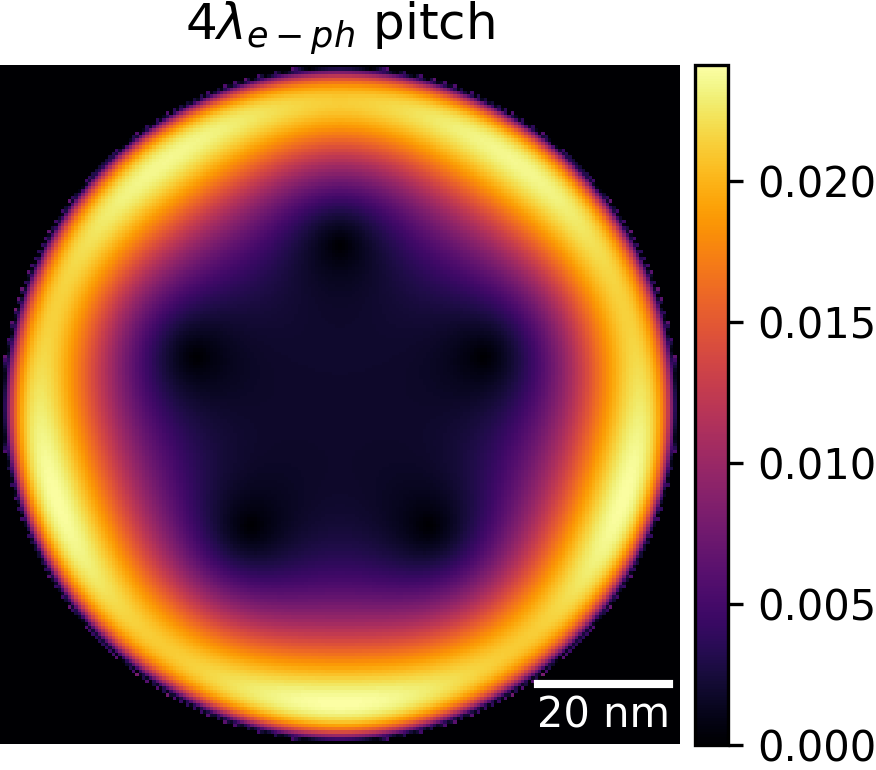}} \\
  \subfloat[]{\label{subfig:model:RCP_4}
  \includegraphics[height=1in]{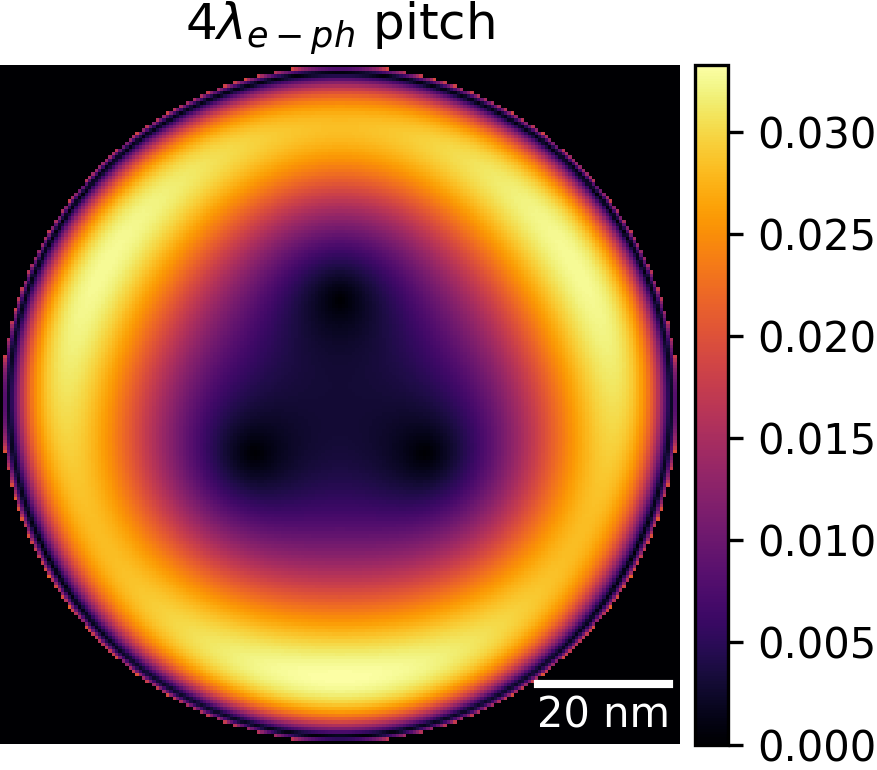}} \\
  \end{minipage}
  \caption{Polarizable 1D helix model predictions for PINEM maps as a function of pitch for (a),(c),(e),(g) LCP and (b),(d),(f),(h) RCP excitation. The spatial pattern is consistent with a superposition of a $0\hbar$ orbital angular momentum mode with a $(\ell+\sigma) \hbar$ orbital angular momentum mode at helical pitches $\ell \lambda_{e-ph}$ with excitation helicity $\sigma$. Calculations correspond to $\unit[100]{\textrm{nm}}$ helix diameter, $\unit[200]{\textrm{keV}}$ electrons and $\unit[800]{\textrm{nm}}$ optical wavelength. \label{fig:model}}
\end{figure}

We numerically integrated the magnetic part of Equation \eqref{eq:dipoles} with a dipole density denfined by this magnetic polarizability with incident circularly polarized light. As we show in Fig. \ref{fig:model}, with a pitch $\ell \lambda_{e-ph}$, where $\lambda_{e-ph} = \frac{2\pi \ve}{\wph}$, and with excitaiton helicity $\sigma$, the coupling constant patterns predicted by this model are superpositions of a small $0\hbar$ and a stronger $(\ell+\sigma )\hbar$ orbital angular momentum mode. This, in combination with the oscillatory behavior of the $0\hbar$ orbital angular momentum mode shown in Fig. 3b in the main text, point to $\lambda_{e-ph}$ as a resonant wavelength for coupling of electrons to the optical field of this structure.

\bibliography{cPINEM_paper,bibliography_supplementals}{}